\documentclass[12pt, twocolumn]{aastex701}

\usepackage{soul}
\usepackage{multirow}
\usepackage{subcaption}

\begin{document}

\title{The Kormendy Relation in the First Billion Years: Evidence from \textit{JWST}}
\shorttitle{The KR at $z\ge$ 6}
\shortauthors{Borgohain \& Saha}

\author[orcid=0000-0002-2870-7716]{Anshuman Borgohain}
\email[show]{anshuman@iucaa.in}
\affiliation{Inter-University Centre for Astronomy and Astrophysics, Pune, Maharashtra - 411007, India
}

\author[orcid=0000-0002-8768-9298]{Kanak Saha}
\email[show]{kanak@iucaa.in}
\affiliation{Inter-University Centre for Astronomy and Astrophysics, Pune, Maharashtra - 411007, India
}

\begin{abstract}
Galaxy scaling relations encode key information about the structural, dynamical, and mass assembly histories of galaxies, and provide constraints on galaxy formation models as well as the onset of galaxy assembly. While these relations are well characterized out to intermediate redshifts, their existence during the first billion years of cosmic history remains largely unconstrained due to observational limitations. In this work, we investigate the Kormendy relation (KR) for spheroidal systems at $z~\ge~6$ using rest-frame $B$-band structural parameters derived from publicly available deep \textit{JWST} imaging of the GOODS, CEERS, PRIMER-UDS, and PRIMER–COSMOS fields.

We find that spheroidal galaxies at these epochs already occupy a well-defined locus in the mean effective surface brightness $(\langle~\mu_{\rm e}~\rangle)$ and effective radius ($\rm~R_{\rm e}$) plane, demonstrating that a KR is already in place when the universe was less a gigayear old. The best-fit relation has a slope of $\beta~=~4.25^{+0.40}_{-0.39}$ and a zero-point of $\alpha~=~15.89^{+0.17}_{-0.17}$, indicating a steeper relation and systematically higher surface brightness compared to the local relation. This steepness reflects the compact sizes and high central stellar-mass densities of these systems, consistent with rapid, dissipative assembly in environments with high gas fractions, likely driven by efficient gas inflows, and gas-rich mergers. The presence of dense bulges embedded in some of these galaxies at similar redshifts further supports a common formation pathway for both bulges and spheroids. Altogether, these findings indicate a predominantly dissipative mode of assembly for the first spheroidal systems which may evolve into the compact quiescent galaxies observed at later cosmic epochs.  
\end{abstract}

\keywords{\uat{Galaxies}{573} --- \uat{Galaxy evolution}{594} --- \uat{Galaxy formation}{595} --- \uat{Galaxy structure}{622} --- \uat{High-redshift galaxies}{734}}

\section{Introduction}
The key observables of galaxies such as luminosity, structure and morphology, size, kinematics, and colour carry crucial information about their formation pathways and evolutionary state. Empirical correlations or scaling relations among these quantities, which are expected to emerge naturally from systems approaching virial equilibrium \citep[e.g.][]{DjorgovskiDavis1987, Dressler1987, Bender1992, Steinmetz_Navarro1999, binney_tremaine2008, DOnofrio_chiosi2022}, serve as stringent constraints on models of galaxy evolution. Classic examples of such scaling relations include the \textit{Tully-Fisher relation} for disk galaxies \citep{TullyFisher1977}, linking luminosity with rotational velocity, and the \textit{Faber-Jackson relation} for spheroidal systems \citep{FaberJackson1976}, relating luminosity to stellar velocity dispersion. A correlation between the effective radius and the mean surface brightness, the \textit{Kormendy relation} (KR), was also identified for early-type systems \citep{Kormendy1977}. These relations were later understood as projections of a more fundamental three-dimensional \textit{Fundamental Plane} connecting the average surface brightness within effective radius, size, and central velocity dispersion \citep{DjorgovskiDavis1987, Dressler1987, Bender1992} for early-type galaxies (ETGs). The slopes and intrinsic scatter of these scaling relations reflect variations in the drivers that regulate the structural and dynamical evolution of galaxies such as dissipative processes, star-formation histories, and environmental effects.

Among the different galaxy scaling relations mentioned above, the KR is particularly valuable and straightforward to examine because it relies solely on photometric observables. It is expressed as

\begin{equation}
    \langle \mu_{\rm e} \rangle = \alpha + \beta \log \mathrm{R}_{\rm e},
\end{equation}

\noindent where $\langle\mu_{\rm e}\rangle$ is the average surface brightness (in mag arcsec$^{-2}$) within the projected half-light radius R$_{\rm e}$ (in kpc), and $\beta$ and $\alpha$ are the slope and zero-point respectively. Extensive work over the past decades has established a well-defined KR for ETGs across the local and intermediate redshift ($z$) Universe \citep[e.g.][]{Ziegler_etal1999, LaBarbera_etal2003, longhetti_etal2007, Rettura_etal2010, saracco_etal2014}. As such, it serves as a powerful diagnostic for investigating the structural and evolutionary pathways of spheroidal systems \citep[e.g.][]{gadotti2009,sachdeva_etal2017,gao_etal2020}, and it holds across a broad range of luminosity and environments \citep[e.g.][]{bernardi_etal2003a, bernardi_etal2003b,LaBarbera_etal2003,LaBarbera_etal2010}. For elliptical galaxies (Es), $\beta$ is typically $\sim 3$ and remains broadly consistent out to $z \sim 1.5$ \citep[e.g.][]{Hamabe_Kormendy1987,longhetti_etal2007,saracco_etal2014,saracco_etal2017,Tortorelli_etal2018, khanday_etal2022}. This relation between effective radius and mean effective surface brightness, with a positive slope in the KR-plane, implies that more luminous elliptical galaxies tend to be larger systems whose light is distributed over a larger physical area, resulting in lower mean effective surface brightnesses.

However, deviations from the KR are also seen that encode the physical processes governing spheroid assembly, thereby reflecting their distinct formation scenarios \citep[e.g.][]{gadotti2009, gao_etal2020, pastrav2021}. Dissipative (wet) mergers or rapid collapse result in strong starbursts and produce compact, high surface-brightness systems, thereby steepening the relation. On the other hand, dissipationless (dry) mergers result in larger and more diffuse spheroids, thereby flattening the slope and increasing the scatter \citep[e.g.][]{bekki1998, nipoti_etal2003, hopkins_etal2009b}. The zero-point of the KR is found to be sensitive to systematic changes in the mass-to-light ratio, stellar population properties, IMF variations, kinematics, and dark matter content \citep[e.g.][]{Prugniel_Simien1996, Forbes_etal1998, Borriello_etal2003, Cappellari_etal2006, auger_etal2010}. Studies further show that the slope and scatter of the relation depend on galaxy mass. This is expected because the KR is a projection of the \textit{Fundamental Plane} along the velocity-dispersion axis. For dispersion-supported systems such as ETGs, the velocity dispersion traces the depth of the gravitational potential well and therefore correlates strongly with galaxy stellar mass. Massive Es are found to follow a tight KR sequence, whereas, low-luminosity systems deviate systematically from this locus, indicating structural non-homology and pointing towards distinct formative and evolutionary pathways \citep{Nigoche-Netro_etal2008,LaBarbera_etal2010}. This is further supported by recent work on extreme low-luminosity systems such as ultra-diffuse galaxies (UDGs) which occupy a different locus in the KR plane \citep{zoller-etal2024}. A similar dichotomy is observed among bulges in disk galaxies, where classical bulges align closely with the ETG KR, indicating similar formation scenarios, whereas pseudo-bulges lie systematically below it \citep{sachdeva_etal2017, sachdeva_etal2020, gao_etal2020}.

Observations at higher redshift provide further insight into the assembly of spheroidal systems at earlier times. \citet{sachdeva_saha2018} discovered a population of bright, compact bulges (0.4 $<z$ $<$ 1.0) with high surface brightness and small radii, suggesting that at least some bulges assembled the bulk of their structure early in cosmic time. At $z \sim $ 1--2, massive ETGs are also observed to be more compact and centrally dense than their local counterparts \citep{longhetti_etal2007, damjanov_etal2009}. Their offsets from the local KR also show that passive luminosity fading alone is insufficient to connect them to low-$z$ counterparts and that substantial size growth is required to bridge the gap \citep[e.g.][]{Rettura_etal2010, van_der_Wel_etal2008,naab_etal2009,saracco_etal2014}. In addition to the above, the measured KR slope also depends on the observed wavelength \citep[e.g.][]{Tortorelli_etal2023} and sample-selection criteria \citep[e.g.][]{saracco_etal2010, Fagioli_etal2016, Tortorelli_etal2018}, highlighting the need for homogeneous analyses when assessing the evolution of the relation over cosmic time. Together, these results highlight that the KR may evolve with redshift, reflecting a possible variation in the relative importance of dissipative process, quenching, and subsequent size growth in the assembly of spheroidal systems.

As highlighted above, studies of the KR at high redshifts have so far been largely restricted up to $z \sim $ 1--2, primarily due to cosmological surface-brightness dimming, limited spatial resolution, and/or the limited coverage of rest-frame optical/near-infrared band imaging at higher redshifts. The advent of \textit{JWST} has transformed this situation with its unprecedented sensitivity and resolution at infrared wavelengths by enabling the detection and structural assessment of compact structures up to $z \sim 11$ in rest-frame optical wavelengths \citep[e.g.][]{ferreira-etal2023, kartaltepe-etal2023, huertas-company-etal2024, baker-etal2025, fujimoto-etal2025, borgohain_saha2025}. This opens, for the first time, the possibility of testing whether the earliest spheroids during the Cosmic Dawn already adhered to a scaling relation such as the KR. This will provide a powerful probe to assess the balance between dissipative collapse, secular evolution, and hierarchical growth, and help identify the progenitors of compact, quiescent galaxies observed at later epochs.

In this work, we investigate the KR for spheroidal systems at $z\ge$ 6 (restricting to galaxies in the first billion years) using rest-frame $B$-band measurements derived from deep \textit{JWST} observations of the Great Observatories Origins Deep Survey (GOODS) fields. We use publicly available morphological catalog for other $JWST$ deep fields--CEERS, PRIMER–UDS, and PRIMER–COSMOS, from DAWN JWST Archive \citep[DJA,][]{genin_etal2025}. Throughout this paper, we adopt a flat $\Lambda$CDM cosmology with $H_{0}=70~\mathrm{km~s^{-1}~Mpc^{-1}}$, $\Omega_{m}=0.3$, and $\Omega_{\Lambda}=0.7$, and all magnitudes are in the AB system \citep{oke1974}.

\section{Dataset}
In this work, we select galaxies that were already assembled by $z=6$, corresponding to $\sim$ $1~\mathrm{Gyr}$ after the $Big~Bang$. The majority of these sources lie in the GOODS-North and GOODS-South fields and have spectroscopically confirmed redshifts ($z_{\rm spec}$) from publicly available \textit{JWST} surveys and catalogs, including JADES \citep{eisenstein-etal2023,eisenstein-etal2023b,rieke-etal2023,hainline-etal2024,bunker-etal2024,deugenio-etal2025} and FRESCO \citep{oesch-etal2023,meyer-etal2024,covelo-paz-etal2025}. The specific observations analyzed can be accessed via the \citeauthor{jades_dataset} (\href{https://doi.org/10.17909/8tdj-8n28}{doi:10.17909/gdyc-7g80}) and \citeauthor{fresco_dataset} (\href{https://doi.org/10.17909/gdyc-7g80}{doi:10.17909/gdyc-7g80}) datasets. To investigate the KR for galaxies in the early Universe ($z\ge6$), we construct a sample of systems exhibiting spheroidal morphology. This sample is drawn from the parent catalog constructed in \citep[][hereafter BS25]{borgohain_saha2025}. The BS25 sample comprises 187 \textit{JWST}-selected galaxies with spectroscopically confirmed redshifts ($z\sim6-11$), for which robust structural parameters were derived from two-dimensional surface-brightness modelling in the restframe optical. We refer the reader to BS25 for a detailed description of the sample selection, structural analysis, and morphological classification procedures.

Morphological measurements for the sources were performed on \textit{JWST} imaging using filters that correspond to the rest-frame $B$ band \textbf{($\sim3000-4000 \AA$)} at the appropriate redshift (BS25). This wavelength regime is chosen to trace rest-frame optical light out to the highest redshifts possible ($\sim6-11$), while capturing emission from young to moderately aged stellar populations that provide a reasonable proxy for the underlying stellar mass distribution at these epochs. 167 out of 187 galaxies in BS25 were modelled using a single Sersic component, and subsequently classified as disk-like or spheroidal systems according to their Sersic indices ($n$). Here, we construct two samples for our study. 

\begin{itemize}
    \item \textbf{Sample 1:} We adopt $n > 1.5$, following \cite{park-etal2022}, as the criterion and select 44 spheroidal systems which is 23\% of the parent sample in BS25. This threshold value of \textit{n} was based on an SDSS sample and provides an effective separation between early- and late-type galaxies.
    \item \textbf{Sample 2:} This sample is based on a more conservative threshold of $n>$ 3 which is commonly used in literature \citep[e.g.][]{Krajnovic_etal2013, hu_etal2024} and therefore serves as a test of the sensitivity of our results to the adopted morphological selection. The sample is essentially a subset of Sample 1 and contain 24 galaxies.
\end{itemize}

\begin{figure*}[htpb]
    \centering
    \includegraphics[width=\linewidth]{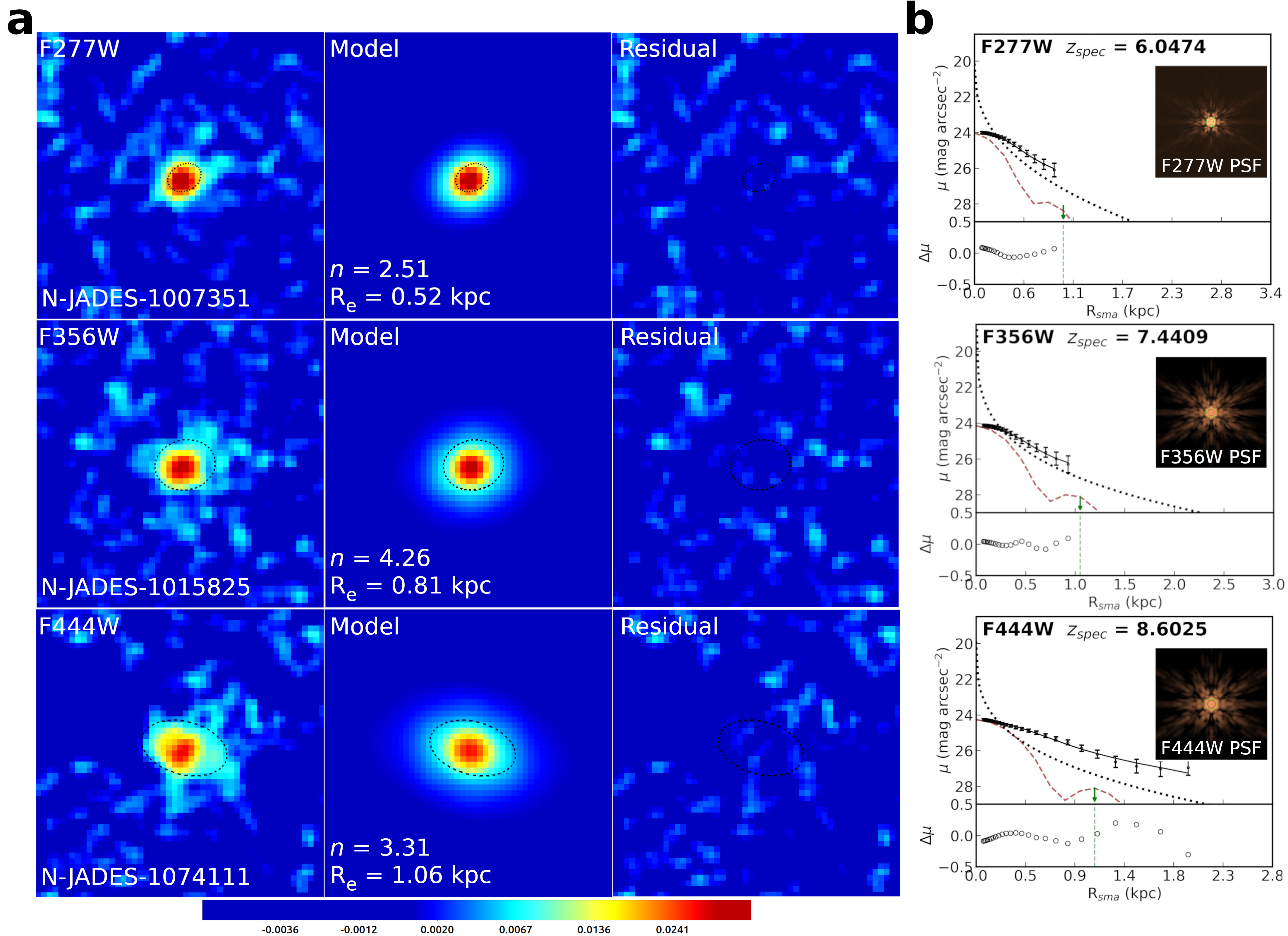}
    \caption{\textbf{(a) False-colour images of a few $z \ge 6$ galaxies from our sample that are best fit with a single Sersic model.} For each galaxy, we present the \textit{JWST} image sampling the rest-frame optical $B$-band, the best-fit model, and the residual image in the first, second, and third column, respectively. The black dotted ellipse represents the effective radius (R$_{\mathrm{e}}$) of the model. \textbf{(b) The 1D surface-brightness profiles of the galaxies.} We present the best-fit surface brightness model profiles (and intrinsic profiles) using solid (dotted) curves. We show the corresponding residual profiles in the bottom panel of each plot. The green inverted arrow marks the 80\% light-enclosing radius of the corresponding \textit{JWST} PSF (shown by the brown dashed curve). We show the PSF of the respective \textit{JWST} filter in the inset, with the 80\% enclosed-flux aperture indicated by a green circle. The spectroscopic redshifts of the galaxies are listed at the top of these panels.}
    \label{fig:galaxy_examples}
\end{figure*}

 The modelling was carried out using the 2D surface-brightness fitting tool \textsc{GALFIT} \citep{peng_etal2002, peng_etal2010} where conservative bounds were imposed on the parameter space to ensure numerical stability and avoid unphysical solutions. The effects of the point-spread function (PSF) were taken into account during the structural modelling. The model PSFs were generated using \textsc{STPSF}/\textsc{WebbPSF} with an oversample factor of 1 and 2 for the short-wavelength (SW) and long-wavelength (LW) filters respectively, to match the pixel scale of the science images (0.03 arcsec/pixel). These PSFs were used to convolve the intrinsic models during the \textsc{GALFIT} modelling. During the modelling, the lower limit on the effective radius R$_{\rm e}$ was set such that it does not fall below the PSF size ($\sim$1.5-pixel HWHM), while the upper limit was fixed at 10 pixels ($0.3''$), corresponding to $\sim 1.7~\mathrm{kpc}$ at $z\sim6$ and $\sim 1~\mathrm{kpc}$ at $z\sim13$. We tested the effect of using these constraints and found no systematic bias in our measurements. We discuss this test in Appendix A of the paper. Additionally, a subset of 20 galaxies in the BS25 sample was successfully modelled with a combination of a Sersic and an Exponential disk -- indicating the presence of a central bulge component embedded within a disk component. For these two-component systems, the Sersic index of the bulge was left free, reflecting the fact that such structures at high redshift are still in the process of assembly. Further details on the modelling procedure can be found in BS25. We present the best-fit 2D \textsc{GALFIT} models and the 1D surface brightness profiles for a few galaxies in Figure \ref{fig:galaxy_examples}, shown in panels (a) and (b), respectively. Based on the centre, position angle and ellipticity of the \textsc{GALFIT} models, we place concentric elliptical isophotes on the parent image and PSF-convolved model image using the \textsc{isophote} module of \textsc{photutils}. The mean intensity along each isophote is then extracted to construct the 1D observed and PSF-convolved model profiles. The intrinsic model profile is computed analytically using the best-fitting structural parameters; taking into account the ellipticity of the model.

Given the limited number of high-$z$ galaxies available in any individual field, we supplement the BS25 sample with $z\ge6$ galaxies from several other deep \textit{JWST} fields as compiled in DJA \citep{genin_etal2025}. This increases the sample size and improves the statistical significance of the structural trends examined in this work. For these additional sources, we adopt the structural parameters directly from the DJA catalog, without re-performing the morphological structure modelling. We exclude objects for which two-component fits were used in the catalog where the Sersic indices for the inner component were fixed to $n=4$ \citep{genin_etal2025}, making them unsuitable for consistent classification within the current framework. Therefore, we obtain 15 (10) galaxies with $n>1.5$ ($n>3$) from the DJA. Our final sample comprises of 59 spheroidal galaxies in Sample 1 (34 in Sample 2) and 20 bulges within disk galaxies at $z\ge6$ with robust structural measurements.

\section{Modelling the high-$z$ KR}
To evaluate the high-$z$ KR, we use the structural parameters obtained from the \textsc{GALFIT} models in BS25. The mean effective surface brightness within the effective radius R$_{\rm e}$ (in arcseconds) is

\begin{equation}
    \langle \mu_{e} \rangle = m + 2.5 \log (2\pi \rm{R}_{\rm e}^{2}),
\end{equation}

where $m$ is the \textsc{GALFIT} model magnitude. We then apply corrections for cosmological surface-brightness dimming, inclination, and foreground extinction as follows:
\begin{equation}
    \langle \mu_{e} \rangle_{\rm corr}
    = \langle \mu_{e} \rangle
    - 10\log(1+z)
    + 2.5\log(q)
    - A_{\lambda},
\end{equation}

where $z$ is the redshift, $q$ the axis ratio, and $A_{\lambda}$ is the foreground extinction in the corresponding $JWST$ filter. The foreground extinction values were obtained from the NASA/IPAC Extragalactic Database (NED) Galactic Dust Reddening and Extinction service, which is based on the dust maps of \citet{Schlegel-etal1998} recalibrated by \citet{SchlaflyFinkbeiner2011}.

We adopt a full Bayesian framework to determine the KR parameters. Unlike traditional approaches that rely on direct estimates obtained through minimization, this method allows us to explore the full posterior distribution of the model parameters. As a result, the uncertainties on the parameters are characterized in a statistically robust manner while naturally accounting for measurement uncertainties and their covariances. This framework also allows the intrinsic scatter of the relation to be treated as a model parameter and thereby it is inferred self-consistently from the data. We perform the analysis using the \texttt{bilby} package \citep{ashton-etal2019}. This approach therefore enables us to assess the stability and robustness of the derived KR parameters and to evaluate the extent to which the relation is constrained by the currently available high-$z$ observations. We define the general form of the likelihood function as:

\begin{equation}
\ln \mathcal{L}
= -\frac{1}{2} \sum_{i}
\left[
\frac{(y_i - m x_i - c)^2}{\sigma_{\mathrm{tot},i}^2}
+ \ln\left(\sigma_{\mathrm{tot},i}^2\right)
\right],
\end{equation}

\noindent where the total variance is given by

\begin{equation}
\sigma_{\mathrm{tot},i}^2
= (m\,\sigma_{x,i})^2 + \sigma_{y,i}^2 + \sigma_{\mathrm{int}}^2.
\end{equation}

In the above, $y_{i}$ corresponds to $\langle \mu_{e} \rangle_{\rm corr}$, $x_{i}$ the observed $\log$R$_{\rm e}$ (in kpc), $\sigma_{x,i}$, $\sigma_{y,i}$ represent their corresponding errors, $\sigma_{\text{int}}$ is the intrinsic scatter and $m$, $c$ corresponds to the slope ($\beta$) and zero-point ($\alpha$) of the KR respectively. 

\section{Results}
\begin{figure*}[htpb]
    \centering
    \includegraphics[width=\linewidth]{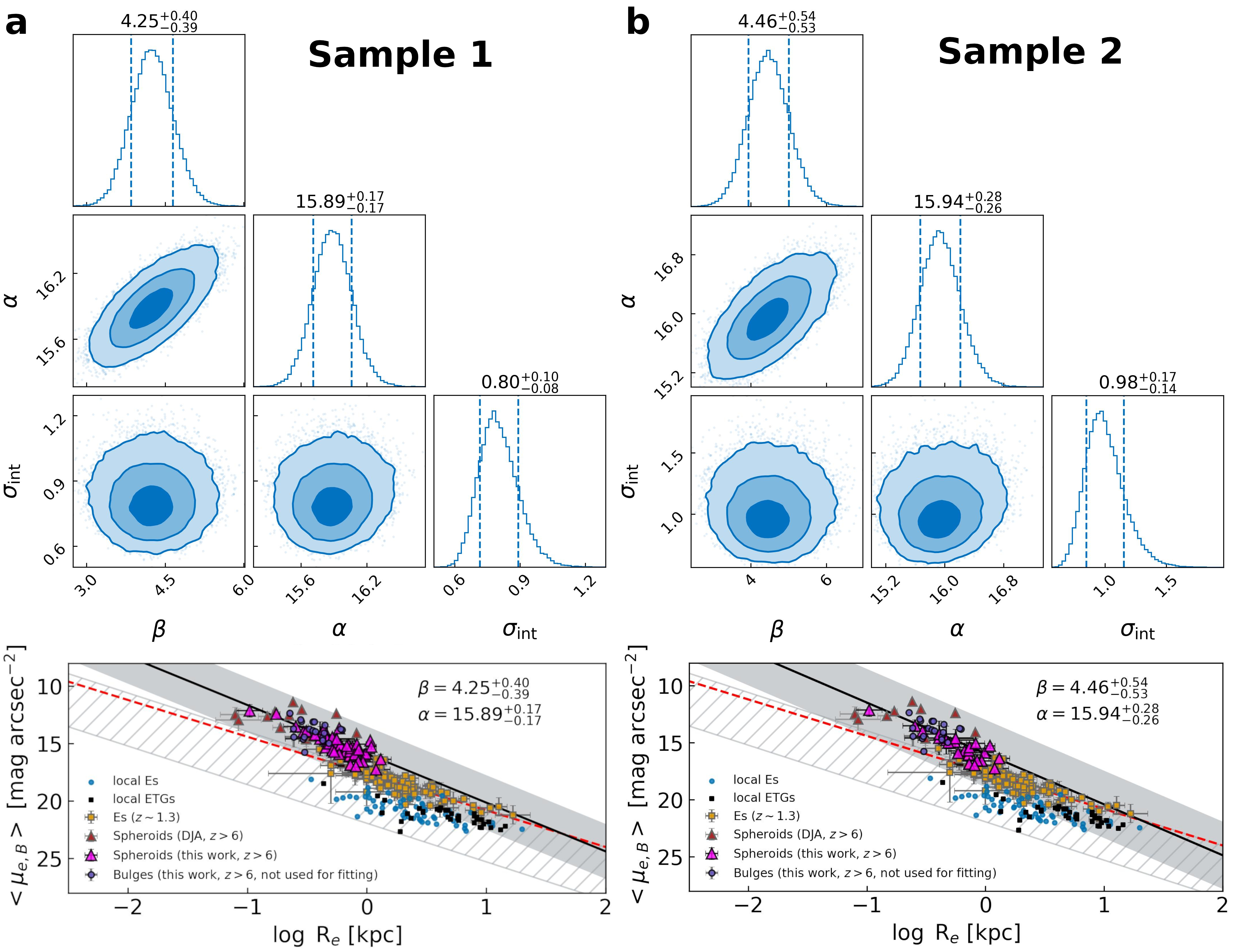}
    \caption{\textbf{(a) Modelling of the KR for Sample 1 ($n>1.5$). Top:} The marginalized posterior distributions for the KR slope, $\beta$, zero-point, $\alpha$ and intrinsic scatter, $\sigma_{int}$. The off-diagonal panels show the 2D joint distributions for the KR parameters, with the contours representing the 1$\sigma$, 2$\sigma$, and 3$\sigma$ probability regions corresponding to a 2D Gaussian distribution. The diagonal panels show the corresponding 1D histograms, with the vertical lines marking the 16$^{\rm{th}}$ and 84$^{\rm{th}}$ percentiles around the median values quoted at the top. \textbf{Bottom:} The black solid curve with gray shaded region represent the best-fit relation with 3$\sigma$ boundaries for spheroidal systems at $z\ge$ 6 in this work. Here, galaxies from other deep fields are also taken into account from the morphological catalog of Dawn \textit{JWST} Archive \citep[DJA,][]{genin_etal2025}. As the bulges in this work have $n<$ 1.5 (BS25), we exclude them from the KR-modelling but place them onto the plot for comparison. The local Es and ETGs are taken from \citet{graham2002,reda_etal2005,khanday_etal2022}. The $z\sim$ 1.3 Es are taken from \citet{saracco_etal2017} and their best-fit relation is given by the red dashed curve. The gray hatched region represent local KR from \citet{khanday_etal2022} with 3$\sigma$ boundaries. \textbf{(b)} Same as panel a, but for Sample 2.}
    \label{fig:KR-combined}
\end{figure*}

In Figure~\ref{fig:KR-combined} (a) and (b), we present the marginalized posterior distributions of the KR parameters obtained from the Bayesian fit using \texttt{bilby} (top panels) and the best-fit KRs (bottom panels) derived from the posterior median values of the model parameters for Sample 1 ($n>1.5$) and Sample 2 ($n>3$), respectively. We obtain best-fit $\beta=$ 4.25$^{+0.40}_{-0.39}$, $\alpha=$ 15.89$^{+0.17}_{-0.17}$ and intrinsic scatter in the relation, $\sigma_{int}=$ 0.80$^{+0.10}_{-0.08}$ for Sample 1. For Sample 2, we find a slope value $\beta=$ 4.46$^{+0.54}_{-0.53}$, zero-point $\alpha=$ 15.94$^{+0.28}_{-0.26}$ and scatter $\sigma_{int}=$ 0.98$^{+0.17}_{-0.14}$. The compact density contours in the posterior distributions indicate that the parameters are well constrained by the data with a mild correlation between $\alpha$ and $\beta$. Additionally, the derived parameters for the two samples are consistent within their errorbars and intrinsic scatter. One might expect a stricter morphological selection to produce a more homogeneous galaxy population and therefore a smaller intrinsic scatter in the KR. However, the intrinsic scatter inferred for the $n>3$ sample remains consistent, within the uncertainties, with that of the broader $n>1.5$ sample. The marginally larger scatter obtained for the $n>3$ sample is likely driven by the smaller sample size and potential incompleteness inherent in our spectroscopically selected sample of high-$z$ spheroidal population. Such consistency of the KR parameters derived from the two samples indicates that our results are not sensitive to the exact choice of the morphological threshold for ETGs. We also note that the methodology employed for deriving the structural parameters in the DJA catalog differs from that used in this work. In contrast to our rest-frame \textit{B}-band measurements, the DJA parameters are derived from multiband modelling and represent weighted averages over the wavelength range 0.8--5 $\mu$m. To assess the impact of these methodological differences, we repeated the KR analysis after excluding the DJA galaxies from our sample. We obtained $\beta=3.81^{+0.50}_{-0.49}$, $\alpha=15.90^{+0.16}_{-0.15}$, and $\sigma_{\rm int}=0.56^{+0.09}_{-0.07}$ (for $n>1.5$ galaxies), which are broadly consistent within the uncertainties with the values derived from the full sample. While the slope and zero-point remain unchanged within their uncertainties, we note that the intrinsic scatter decreases when the DJA galaxies are excluded. This indicates a possibility that the DJA galaxies may be contributing to a diversity in the high-$z$ spheroidal population in the KR plane. However, the current sample is insufficient to assess whether this decrease in scatter reflects a genuine variation in the population or simply arises from sample variance. The KR measurements for both the $n>1.5$ and $n>3$ samples, excluding the DJA galaxies, are presented in Appendix B. Note that we exclude bulge components at $z\ge6$ from the KR modelling because they have Sersic indices $n<1.5$ (BS25).

In the bottom panels of Figure~\ref{fig:KR-combined}, we present a comparison of our morphologically chosen ETG sample at $z\ge6$ with the local Es and ETGs from \citet{graham2002,reda_etal2005,khanday_etal2022}, and intermediate redshift ($z\sim$1.3) Es from \citet{saracco_etal2017} in the $\langle \mu_{\rm e} \rangle-\rm R_{\rm e}$ plane. This highlights the compact nature and higher surface brightness of ETGs during the first billion years. We also compare the KRs reported from recent literature \citep{saracco_etal2017, khanday_etal2022} with our derived relation. In the local Universe, \citet{khanday_etal2022} report a slope of $\beta \sim3$ for ETGs and dwarf Es in the $g$-band, while \citet{saracco_etal2017} find a similar slope of $\beta \sim 3$ in the rest-frame $B$-band for Es at $z \sim 1.3$. This indicates that, whereas previous studies have found little evolution in redshift in the KR slope from $z\sim1.3$ to the local Universe ($\sim9$ billion years), our derived steeper relation indicates evolution at earlier epochs. This implies that a considerable fraction of the structural growth of Es may occur rapidly within the first few billion years of cosmic history and followed by a much slower evolution thereafter. It should be noted that $\beta$ in the bluer bands may be affected by strong stellar population gradients \citep[e.g.,][]{Tortorelli_etal2023}. For instance, the best-fit \textit{Fundamental Plane} in \citet{bernardi_etal2003b} corresponds a slope of $\beta \sim 3.3$ in the $g^{*}$-band, while \citet{LaBarbera_etal2010} report a $g$-band $\beta ~\text{value} \sim 3.44$. \citet{Tortorelli_etal2023} report a broader range of slopes, varying from $\beta \sim 2.7$ to $\sim 3.86$ in the rest-frame $B$ band. These variations highlight that the measured KR slope in bluer bands can be sensitive to stellar population gradients within galaxies. However, in the case of galaxies at $z\ge6$, with only a $\sim$Gyr timescale of evolution, such gradients are expected to be minimal.

\begin{figure}[htpb]
\centering
\includegraphics[width=1.05\linewidth]{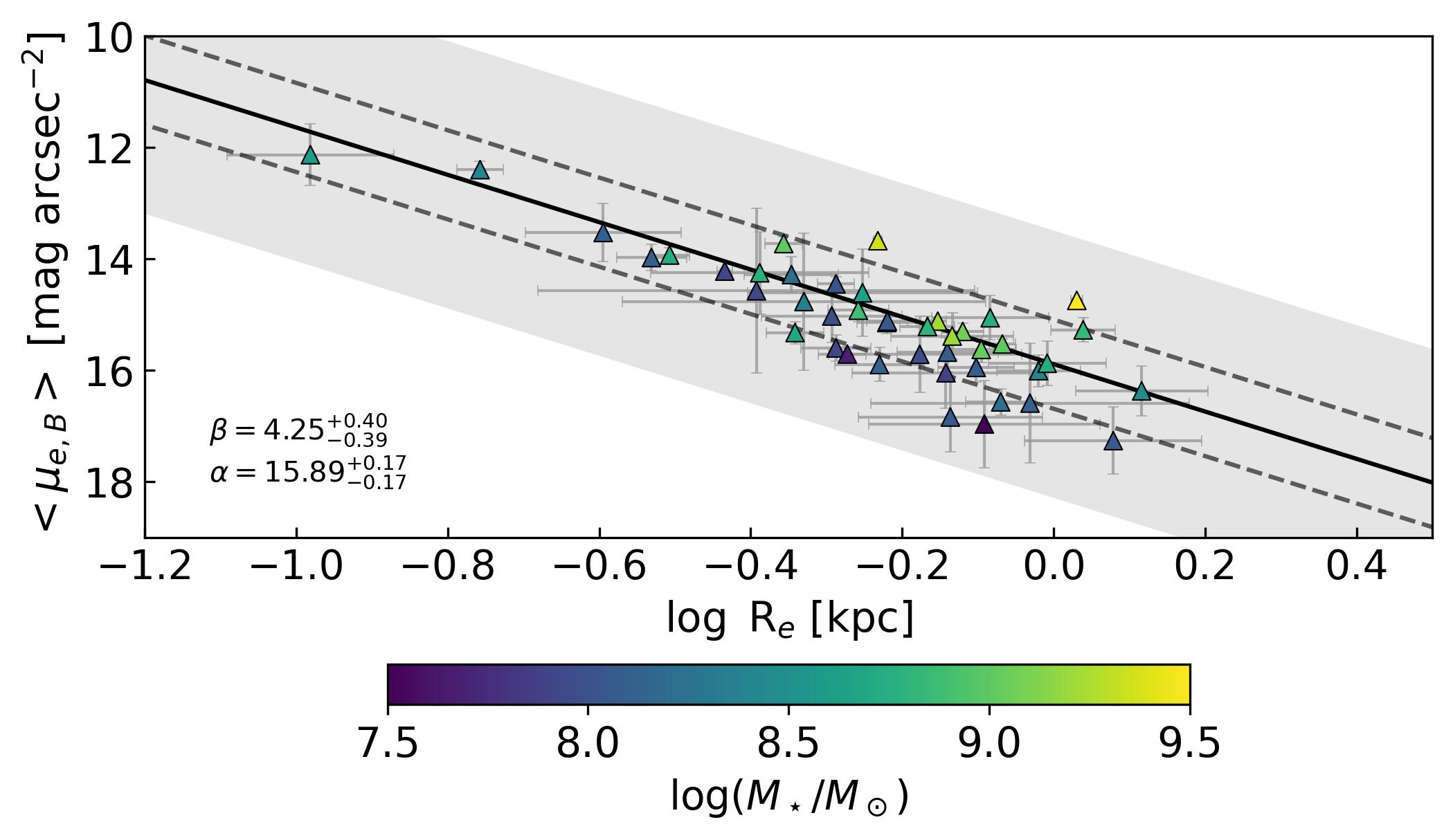}
\caption{\textbf{The Kormendy relation for the Sample~1 ($n>1.5$) galaxies colour-coded by stellar mass.} The solid black line shows the best-fitting relation, while the dashed and shaded regions denote the corresponding 1$\sigma$ and 3$\sigma$ scatter, respectively. The errorbars denote 1$\sigma$ errors.}
\label{fig:KR-mass}
\end{figure}

We explore whether stellar mass influences the location of galaxies on the KR and, consequently, whether a \textit{Fundamental Plane} may already be in place at these early epochs. Figure~\ref{fig:KR-mass} shows the KR for our Sample~1 galaxies colour-coded by stellar mass. While lower-mass galaxies appear to occupy the fainter side of the relation, we do not find clear evidence for a robust segregation of galaxies by stellar mass, as systems of different masses largely overlap in the KR plane. This may partly be due to the relatively narrow stellar-mass range spanned by the current sample, (7.5 $\lesssim \log$(M$_\star$/M$_\odot$) $\lesssim$ 9.5), which is dominated by low-mass galaxies. Consequently, the present data do not allow a strong test of whether a \textit{Fundamental Plane} is already established at these redshifts. Future studies based on larger samples and spanning a broader stellar-mass range will provide a stronger test of this possibility. Direct velocity-dispersion measurements from \textit{JWST} spectroscopy will also be crucial for establishing whether a \textit{Fundamental Plane} is already in place within the first billion years of cosmic history.

\section{Discussion}
Our results based on deep $JWST$ observations suggest that the scaling relation for spheroidal systems may already be in place at very early cosmic epochs. As presented in the previous section, these young systems at $z\ge6$ follow a steeper KR as compared to what has been observed in the local and intermediate-$z$ Universe. In a simple scenario, the effective radius $\text{R}_{e}\propto \text{I}_{e}^{-1}$ for a virialised system (under dissipationless limit), where $\text{I}_{e}$ is the effective surface brightness of the system. This translates to a value of $\beta= 2.5$. However, the dissipation limit for a collapsing system implies $\text{R}_{e}\propto \text{I}_{e}^{-0.5}$, which gives a value of $\beta=5$. Our findings of $\beta=4.25$ for the $z\ge6$ sample of spheroidal systems have important implications for understanding the assembly and long-term structural evolution of such systems. A steeper-than-local relation means that, at fixed R$_{\rm{e}}$, these systems are more luminous and denser than their local counterparts - an indication of a possible enhancement in the role of dissipative processes in their assembly. Galaxies at these epochs are expected to exhibit high and evolving gas-fractions \citep[e.g.][]{dekel_etal2009,tacconi_etal2010, Tacconi-etal2020, Tacchella-etal2023}, wet-mergers \citep{robertson_etal2006, hopkins_etal2008, hopkins_etal2009a, puskas-etal2025, hu-etal2025, duan-etal2025a, duan-etal2026} and efficient cold gas inflows and star-formation \citep[e.g.,][]{Heintz-etal2023,Pollock-etal2026}. Together, these processes drive rapid, centrally concentrated star formation and compaction \citep{ji-etal2024, baker-etal2025}, naturally producing the observed compact systems with high central stellar mass densities \citep{baggen-etal2023, morishita-etal2024} and systematically place these systems above the local KR. Theoretical models further suggest that such systems may also form through `$quasi-monolithic$' collapse, wherein gas-clouds assemble rapidly in a hierarchical fashion and form stars in an intense burst \citep[e.g.][]{merlin_etal2012} - similar to `blue-nuggets' \citep{dekel_burkert2014,zolotov-etal2015}. It is quite possible that these systems are progenitors of massive quiescent systems at later times \citep{carnall_etal2023,deGraff_etal2025}. Subsequent evolution may also proceed via passive luminosity fading and/or dissipationless mergers that will drive the systems towards the present-day KR \citep[e.g.][]{hopkins_etal2009a, naab_etal2009}. In fact, the current picture is naturally consistent with the two-phase assembly framework for such systems \citep[e.g.,][]{naab_etal2007, oser_etal2010}. The compact, high-surface-brightness systems that we observe at $z\ge6$ represent the products of the first, predominantly dissipative phase involving rapid in-situ star formation, driven by cold gas inflows and gas-rich mergers that dominate until $z\sim2$ \citep{oser_etal2010}. These systems then grow through an extended phase of dissipationless accretion of smaller satellite systems, assimilating stellar mass preferentially at large radii, reducing the effective density and driving the relation toward a shallower slope at later times. Tracing the KR across cosmic time, from the compact high-$z$ locus to the local relation, will be essential to quantifying this structural transformation and testing two-phase assembly models. Interestingly, although the bulges in our sample (BS25) were not included in the KR fitting itself, their location in the KR points to a possible structural continuity between early bulges and spheroids, both likely rooted in rapid dissipative assembly. However, such a claim is uncertain and beyond the scope of this work, and will be addressed with larger samples and homogeneous structural measurements in the future.

\section{Conclusions}
We investigate the KR for spheroidal systems at $z\ge6$, using measurements of the effective radius and mean effective surface brightness in the rest-frame $B$-band. The structural parameters are derived from deep \textit{JWST} imaging across the GOODS, CEERS, PRIMER-UDS, and PRIMER-COSMOS survey fields. Our analysis provides one of the earliest direct observational constraints on a fundamental galaxy scaling relation during the epoch of Cosmic Dawn. Our main conclusions are as follows:

\begin{itemize}
    \item Spheroidal systems at $z\ge$ 6 follow a steeper KR with compact sizes and high surface brightness compared to their lower-$z$ counterparts. This indicates that the fundamental galaxy scaling relations are already emerging/in-place within the first Gyr after the $Big~Bang$.
    
    \item The steepness of the relation is consistent with a rapid, dissipative assembly of galaxies or their cores at these epochs, likely driven by high gas fractions and efficient central mass buildup. These systems may possibly evolve into massive quiescent galaxies discovered at $z\sim$ 4--5 or into present-day systems via dissipationless size growth or secular evolution.
\end{itemize}

Future work combining larger samples across cosmic time with spatially resolved stellar population constraints will be essential to determine whether a continuous structural evolutionary sequence links early spheroids, dense bulges in disks, and present-day spheroidal systems. 

\section*{Acknowledements}
This work is based on observations taken by the 3D-HST Treasury Program (GO 12177 and 12328) with the NASA/ESA HST, which is operated by the Association of Universities for Research in Astronomy, Inc., under NASA contract NAS5-26555. This work is based on observations made with the NASA/ESA/CSA James Webb Space Telescope. The data were obtained from the Mikulski Archive for Space Telescopes at the Space Telescope Science Institute, which is operated by the Association of Universities for Research in Astronomy, Inc., under NASA contract NAS 5-03127 for \textit{JWST}. These observations are associated with programmes 1180, 1181, 1210, 1286, 1287, 1895, 1963, and 3215.

\bibliography{references}

\begin{thebibliography}{}
\expandafter\ifx\csname natexlab\endcsname\relax\def\natexlab#1{#1}\fi
\providecommand{\url}[1]{\href{#1}{#1}}
\providecommand{\dodoi}[1]{doi:~\href{http://doi.org/#1}{\nolinkurl{#1}}}
\providecommand{\doeprint}[1]{\href{http://ascl.net/#1}{\nolinkurl{http://ascl.net/#1}}}
\providecommand{\doarXiv}[1]{\href{https://arxiv.org/abs/#1}{\nolinkurl{https://arxiv.org/abs/#1}}}

\bibitem[{{Ashton} {et~al.}(2019){Ashton}, {H{\"u}bner}, {Lasky}, {Talbot},
  {Ackley}, {Biscoveanu}, {Chu}, {Divakarla}, {Easter}, {Goncharov}, {Hernandez
  Vivanco}, {Harms}, {Lower}, {Meadors}, {Melchor}, {Payne}, {Pitkin},
  {Powell}, {Sarin}, {Smith}, \& {Thrane}}]{ashton-etal2019}
{Ashton}, G., {H{\"u}bner}, M., {Lasky}, P.~D., {et~al.} 2019, \apjs, 241, 27,
  \dodoi{10.3847/1538-4365/ab06fc}

\bibitem[{{Auger} {et~al.}(2010){Auger}, {Treu}, {Bolton}, {Gavazzi},
  {Koopmans}, {Marshall}, {Moustakas}, \& {Burles}}]{auger_etal2010}
{Auger}, M.~W., {Treu}, T., {Bolton}, A.~S., {et~al.} 2010, \apj, 724, 511,
  \dodoi{10.1088/0004-637X/724/1/511}

\bibitem[{{Baggen} {et~al.}(2023){Baggen}, {van Dokkum}, {Labb{\'e}},
  {Brammer}, {Miller}, {Bezanson}, {Leja}, {Wang}, {Whitaker}, {Suess}, \&
  {Nelson}}]{baggen-etal2023}
{Baggen}, J. F.~W., {van Dokkum}, P., {Labb{\'e}}, I., {et~al.} 2023, \apjl,
  955, L12, \dodoi{10.3847/2041-8213/acf5ef}

\bibitem[{{Baker} {et~al.}(2025){Baker}, {Tacchella}, {Johnson}, {Nelson},
  {Suess}, {D'Eugenio}, {Curti}, {de Graaff}, {Ji}, {Maiolino}, {Robertson},
  {Scholtz}, {Alberts}, {Arribas}, {Boyett}, {Bunker}, {Carniani}, {Charlot},
  {Chen}, {Chevallard}, {Curtis-Lake}, {Danhaive}, {DeCoursey}, {Egami},
  {Eisenstein}, {Endsley}, {Hausen}, {Helton}, {Kumari}, {Looser}, {Maseda},
  {Pusk{\'a}s}, {Rieke}, {Sandles}, {Sun}, {{\"U}bler}, {Williams}, {Willmer},
  \& {Witstok}}]{baker-etal2025}
{Baker}, W.~M., {Tacchella}, S., {Johnson}, B.~D., {et~al.} 2025, Nature
  Astronomy, 9, 141, \dodoi{10.1038/s41550-024-02384-8}

\bibitem[{{Bekki}(1998)}]{bekki1998}
{Bekki}, K. 1998, \apj, 496, 713, \dodoi{10.1086/305411}

\bibitem[{{Bender} {et~al.}(1992){Bender}, {Burstein}, \& {Faber}}]{Bender1992}
{Bender}, R., {Burstein}, D., \& {Faber}, S.~M. 1992, \apj, 399, 462,
  \dodoi{10.1086/171940}

\bibitem[{{Bernardi} {et~al.}(2003{\natexlab{a}}){Bernardi}, {Sheth}, {Annis},
  {Burles}, {Eisenstein}, {Finkbeiner}, {Hogg}, {Lupton}, {Schlegel},
  {SubbaRao}, {Bahcall}, {Blakeslee}, {Brinkmann}, {Castander}, {Connolly},
  {Csabai}, {Doi}, {Fukugita}, {Frieman}, {Heckman}, {Hennessy}, {Ivezi{\'c}},
  {Knapp}, {Lamb}, {McKay}, {Munn}, {Nichol}, {Okamura}, {Schneider}, {Thakar},
  \& {York}}]{bernardi_etal2003a}
{Bernardi}, M., {Sheth}, R.~K., {Annis}, J., {et~al.} 2003{\natexlab{a}}, \aj,
  125, 1849, \dodoi{10.1086/374256}

\bibitem[{{Bernardi} {et~al.}(2003{\natexlab{b}}){Bernardi}, {Sheth}, {Annis},
  {Burles}, {Eisenstein}, {Finkbeiner}, {Hogg}, {Lupton}, {Schlegel},
  {SubbaRao}, {Bahcall}, {Blakeslee}, {Brinkmann}, {Castander}, {Connolly},
  {Csabai}, {Doi}, {Fukugita}, {Frieman}, {Heckman}, {Hennessy}, {Ivezi{\'c}},
  {Knapp}, {Lamb}, {McKay}, {Munn}, {Nichol}, {Okamura}, {Schneider}, {Thakar},
  \& {York}}]{bernardi_etal2003b}
---. 2003{\natexlab{b}}, \aj, 125, 1866, \dodoi{10.1086/367794}

\bibitem[{{Binney} \& {Tremaine}(2008)}]{binney_tremaine2008}
{Binney}, J., \& {Tremaine}, S. 2008, {Galactic Dynamics: Second Edition}

\bibitem[{{Borgohain} \& {Saha}(2025)}]{borgohain_saha2025}
{Borgohain}, A., \& {Saha}, K. 2025, arXiv e-prints, arXiv:2510.25383,
  \dodoi{10.48550/arXiv.2510.25383}

\bibitem[{{Borriello} {et~al.}(2003){Borriello}, {Salucci}, \&
  {Danese}}]{Borriello_etal2003}
{Borriello}, A., {Salucci}, P., \& {Danese}, L. 2003, \mnras, 341, 1109,
  \dodoi{10.1046/j.1365-8711.2003.06404.x}

\bibitem[{{Bunker} {et~al.}(2024){Bunker}, {Cameron}, {Curtis-Lake},
  {Jakobsen}, {Carniani}, {Curti}, {Witstok}, {Maiolino}, {D'Eugenio},
  {Looser}, {Willott}, {Bonaventura}, {Hainline}, {{\"U}bler}, {Willmer},
  {Saxena}, {Smit}, {Alberts}, {Arribas}, {Baker}, {Baum}, {Bhatawdekar},
  {Bowler}, {Boyett}, {Charlot}, {Chen}, {Chevallard}, {Circosta}, {DeCoursey},
  {de Graaff}, {Egami}, {Eisenstein}, {Endsley}, {Ferruit}, {Giardino},
  {Hausen}, {Helton}, {Hviding}, {Ji}, {Johnson}, {Jones}, {Kumari}, {Laseter},
  {L{\"u}tzgendorf}, {Maseda}, {Nelson}, {Parlanti}, {Perna}, {Rauscher},
  {Rawle}, {Rix}, {Rieke}, {Robertson}, {Rodr{\'\i}guez Del Pino}, {Sandles},
  {Scholtz}, {Sharpe}, {Skarbinski}, {Stark}, {Sun}, {Tacchella}, {Topping},
  {Villanueva}, {Wallace}, {Williams}, \& {Woodrum}}]{bunker-etal2024}
{Bunker}, A.~J., {Cameron}, A.~J., {Curtis-Lake}, E., {et~al.} 2024, \aap, 690,
  A288, \dodoi{10.1051/0004-6361/202347094}

\bibitem[{{Cappellari} {et~al.}(2006){Cappellari}, {Bacon}, {Bureau}, {Damen},
  {Davies}, {de Zeeuw}, {Emsellem}, {Falc{\'o}n-Barroso}, {Krajnovi{\'c}},
  {Kuntschner}, {McDermid}, {Peletier}, {Sarzi}, {van den Bosch}, \& {van de
  Ven}}]{Cappellari_etal2006}
{Cappellari}, M., {Bacon}, R., {Bureau}, M., {et~al.} 2006, \mnras, 366, 1126,
  \dodoi{10.1111/j.1365-2966.2005.09981.x}

\bibitem[{{Carnall} {et~al.}(2023){Carnall}, {McLure}, {Dunlop}, {McLeod},
  {Wild}, {Cullen}, {Magee}, {Begley}, {Cimatti}, {Donnan}, {Hamadouche},
  {Jewell}, \& {Walker}}]{carnall_etal2023}
{Carnall}, A.~C., {McLure}, R.~J., {Dunlop}, J.~S., {et~al.} 2023, \nat, 619,
  716, \dodoi{10.1038/s41586-023-06158-6}

\bibitem[{{Covelo-Paz} {et~al.}(2025){Covelo-Paz}, {Giovinazzo}, {Oesch},
  {Meyer}, {Weibel}, {Brammer}, {Fudamoto}, {Kerutt}, {Lin}, {Matharu},
  {Naidu}, {Velichko}, {Bollo}, {Bouwens}, {Chisholm}, {Illingworth},
  {Kramarenko}, {Magee}, {Maseda}, {Matthee}, {Nelson}, {Reddy}, {Schaerer},
  {Stefanon}, \& {Xiao}}]{covelo-paz-etal2025}
{Covelo-Paz}, A., {Giovinazzo}, E., {Oesch}, P.~A., {et~al.} 2025, \aap, 694,
  A178, \dodoi{10.1051/0004-6361/202452363}

\bibitem[{{Damjanov} {et~al.}(2009){Damjanov}, {McCarthy}, {Abraham},
  {Glazebrook}, {Yan}, {Mentuch}, {Le Borgne}, {Savaglio}, {Crampton},
  {Murowinski}, {Juneau}, {Carlberg}, {J{\o}rgensen}, {Roth}, {Chen}, \&
  {Marzke}}]{damjanov_etal2009}
{Damjanov}, I., {McCarthy}, P.~J., {Abraham}, R.~G., {et~al.} 2009, \apj, 695,
  101, \dodoi{10.1088/0004-637X/695/1/101}

\bibitem[{{de Graaff} {et~al.}(2025){de Graaff}, {Setton}, {Brammer}, {Cutler},
  {Suess}, {Labb{\'e}}, {Leja}, {Weibel}, {Maseda}, {Whitaker}, {Bezanson},
  {Boogaard}, {Cleri}, {De Lucia}, {Franx}, {Greene}, {Hirschmann}, {Matthee},
  {McConachie}, {Naidu}, {Oesch}, {Price}, {Rix}, {Valentino}, {Wang}, \&
  {Williams}}]{deGraff_etal2025}
{de Graaff}, A., {Setton}, D.~J., {Brammer}, G., {et~al.} 2025, Nature
  Astronomy, 9, 280, \dodoi{10.1038/s41550-024-02424-3}

\bibitem[{{Dekel} \& {Burkert}(2014)}]{dekel_burkert2014}
{Dekel}, A., \& {Burkert}, A. 2014, \mnras, 438, 1870,
  \dodoi{10.1093/mnras/stt2331}

\bibitem[{{Dekel} {et~al.}(2009){Dekel}, {Sari}, \&
  {Ceverino}}]{dekel_etal2009}
{Dekel}, A., {Sari}, R., \& {Ceverino}, D. 2009, \apj, 703, 785,
  \dodoi{10.1088/0004-637X/703/1/785}

\bibitem[{{D'Eugenio} {et~al.}(2025){D'Eugenio}, {Cameron}, {Scholtz},
  {Carniani}, {Willott}, {Curtis-Lake}, {Bunker}, {Parlanti}, {Maiolino},
  {Willmer}, {Jakobsen}, {Robertson}, {Johnson}, {Tacchella}, {Cargile},
  {Rawle}, {Arribas}, {Chevallard}, {Curti}, {Egami}, {Eisenstein}, {Kumari},
  {Looser}, {Rieke}, {Rodr{\'\i}guez Del Pino}, {Saxena}, {{\"U}bler},
  {Venturi}, {Witstok}, {Baker}, {Bhatawdekar}, {Bonaventura}, {Boyett},
  {Charlot}, {Danhaive}, {Hainline}, {Hausen}, {Helton}, {Ji}, {Ji}, {Jones},
  {Juod{\v{z}}balis}, {Maseda}, {P{\'e}rez-Gonz{\'a}lez}, {Perna},
  {Pusk{\'a}s}, {Shivaei}, {Silcock}, {Simmonds}, {Smit}, {Sun}, {Villanueva},
  {Williams}, \& {Zhu}}]{deugenio-etal2025}
{D'Eugenio}, F., {Cameron}, A.~J., {Scholtz}, J., {et~al.} 2025, \apjs, 277, 4,
  \dodoi{10.3847/1538-4365/ada148}

\bibitem[{{Djorgovski} \& {Davis}(1987)}]{DjorgovskiDavis1987}
{Djorgovski}, S., \& {Davis}, M. 1987, \apj, 313, 59, \dodoi{10.1086/164948}

\bibitem[{{Dressler} {et~al.}(1987){Dressler}, {Lynden-Bell}, {Burstein},
  {Davies}, {Faber}, {Terlevich}, \& {Wegner}}]{Dressler1987}
{Dressler}, A., {Lynden-Bell}, D., {Burstein}, D., {et~al.} 1987, \apj, 313,
  42, \dodoi{10.1086/164947}

\bibitem[{{Duan} {et~al.}(2025){Duan}, {Conselice}, {Li}, {Austin}, {Harvey},
  {Adams}, {Duncan}, {Trussler}, {Ferreira}, {Westcott}, {Harris}, {Windhorst},
  {Holwerda}, {Broadhurst}, {Coe}, {Cohen}, {Du}, {Driver}, {Frye}, {Grogin},
  {Hathi}, {Jansen}, {Koekemoer}, {Marshall}, {Nonino}, {Ortiz}, {Pirzkal},
  {Robotham}, {Ryan}, {Summers}, {D'Silva}, {Willmer}, \&
  {Yan}}]{duan-etal2025a}
{Duan}, Q., {Conselice}, C.~J., {Li}, Q., {et~al.} 2025, \mnras, 540, 774,
  \dodoi{10.1093/mnras/staf638}

\bibitem[{{Duan} {et~al.}(2026){Duan}, {Conselice}, {Harvey}, {Li}, {Austin},
  {Adams}, {Ferreira}, {Duncan}, {Trussler}, {Pascalau}, {Windhorst},
  {Holwerda}, {Broadhurst}, {Coe}, {Cohen}, {Du}, {Driver}, {Frye}, {Grogin},
  {Hathi}, {Jansen}, {Koekemoer}, {Marshall}, {Nonino}, {Ortiz}, {Pirzkal},
  {Robotham}, {Ryan}, {Summers}, {D'Silva}, {Willmer}, \&
  {Yan}}]{duan-etal2026}
{Duan}, Q., {Conselice}, C.~J., {Harvey}, T., {et~al.} 2026, \mnras, 546,
  stag008, \dodoi{10.1093/mnras/stag008}

\bibitem[{D’Onofrio \& Chiosi(2022)}]{DOnofrio_chiosi2022}
D’Onofrio, M., \& Chiosi, C. 2022, Universe, 8,
  \dodoi{10.3390/universe8010008}

\bibitem[{{Eisenstein} {et~al.}(2023{\natexlab{a}}){Eisenstein}, {Willott},
  {Alberts}, {Arribas}, {Bonaventura}, {Bunker}, {Cameron}, {Carniani},
  {Charlot}, {Curtis-Lake}, {D'Eugenio}, {Endsley}, {Ferruit}, {Giardino},
  {Hainline}, {Hausen}, {Jakobsen}, {Johnson}, {Maiolino}, {Rieke}, {Rieke},
  {Rix}, {Robertson}, {Stark}, {Tacchella}, {Williams}, {Willmer}, {Baker},
  {Baum}, {Bhatawdekar}, {Boyett}, {Chen}, {Chevallard}, {Circosta}, {Curti},
  {Danhaive}, {DeCoursey}, {de Graaff}, {Dressler}, {Egami}, {Helton},
  {Hviding}, {Ji}, {Jones}, {Kumari}, {L{\"u}tzgendorf}, {Laseter}, {Looser},
  {Lyu}, {Maseda}, {Nelson}, {Parlanti}, {Perna}, {Pusk{\'a}s}, {Rawle},
  {Rodr{\'\i}guez Del Pino}, {Sandles}, {Saxena}, {Scholtz}, {Sharpe},
  {Shivaei}, {Silcock}, {Simmonds}, {Skarbinski}, {Smit}, {Stone}, {Suess},
  {Sun}, {Tang}, {Topping}, {{\"U}bler}, {Villanueva}, {Wallace}, {Whitler},
  {Witstok}, \& {Woodrum}}]{eisenstein-etal2023}
{Eisenstein}, D.~J., {Willott}, C., {Alberts}, S., {et~al.} 2023{\natexlab{a}},
  arXiv e-prints, arXiv:2306.02465, \dodoi{10.48550/arXiv.2306.02465}

\bibitem[{{Eisenstein} {et~al.}(2023{\natexlab{b}}){Eisenstein}, {Johnson},
  {Robertson}, {Tacchella}, {Hainline}, {Jakobsen}, {Maiolino}, {Bonaventura},
  {Bunker}, {Cameron}, {Cargile}, {Curtis-Lake}, {Hausen}, {Pusk{\'a}s},
  {Rieke}, {Sun}, {Willmer}, {Willott}, {Alberts}, {Arribas}, {Baker}, {Baum},
  {Bhatawdekar}, {Carniani}, {Charlot}, {Chen}, {Chevallard}, {Curti},
  {DeCoursey}, {D'Eugenio}, {de Graaff}, {Egami}, {Helton}, {Ji}, {Jones},
  {Kumari}, {L{\"u}tzgendorf}, {Laseter}, {Looser}, {Lyu}, {Maseda}, {Nelson},
  {Parlanti}, {Rauscher}, {Rawle}, {Rieke}, {Rix}, {Rujopakarn}, {Sandles},
  {Saxena}, {Scholtz}, {Sharpe}, {Shivaei}, {Simmonds}, {Smit}, {Topping},
  {{\"U}bler}, {Venturi}, {Williams}, {Witstok}, \&
  {Woodrum}}]{eisenstein-etal2023b}
{Eisenstein}, D.~J., {Johnson}, B.~D., {Robertson}, B., {et~al.}
  2023{\natexlab{b}}, arXiv e-prints, arXiv:2310.12340,
  \dodoi{10.48550/arXiv.2310.12340}

\bibitem[{{Faber} \& {Jackson}(1976)}]{FaberJackson1976}
{Faber}, S.~M., \& {Jackson}, R.~E. 1976, \apj, 204, 668,
  \dodoi{10.1086/154215}

\bibitem[{{Fagioli} {et~al.}(2016){Fagioli}, {Carollo}, {Renzini}, {Lilly},
  {Onodera}, \& {Tacchella}}]{Fagioli_etal2016}
{Fagioli}, M., {Carollo}, C.~M., {Renzini}, A., {et~al.} 2016, \apj, 831, 173,
  \dodoi{10.3847/0004-637X/831/2/173}

\bibitem[{{Ferreira} {et~al.}(2023){Ferreira}, {Conselice}, {Sazonova},
  {Ferrari}, {Caruana}, {Tohill}, {Lucatelli}, {Adams}, {Irodotou}, {Marshall},
  {Roper}, {Lovell}, {Verma}, {Austin}, {Trussler}, \&
  {Wilkins}}]{ferreira-etal2023}
{Ferreira}, L., {Conselice}, C.~J., {Sazonova}, E., {et~al.} 2023, \apj, 955,
  94, \dodoi{10.3847/1538-4357/acec76}

\bibitem[{{Forbes} {et~al.}(1998){Forbes}, {Ponman}, \&
  {Brown}}]{Forbes_etal1998}
{Forbes}, D.~A., {Ponman}, T.~J., \& {Brown}, R. J.~N. 1998, \apjl, 508, L43,
  \dodoi{10.1086/311715}

\bibitem[{{FRESCO}(2023)}]{fresco_dataset}
{FRESCO}. 2023, The JWST FRESCO Survey,  STScI/MAST, \dodoi{10.17909/GDYC-7G80}

\bibitem[{{Fujimoto} {et~al.}(2025){Fujimoto}, {Ouchi}, {Kohno}, {Valentino},
  {Gim{\'e}nez-Arteaga}, {Brammer}, {Furtak}, {Kohandel}, {Oguri},
  {Pallottini}, {Richard}, {Zitrin}, {Bauer}, {Boylan-Kolchin},
  {Dessauges-Zavadsky}, {Egami}, {Finkelstein}, {Ma}, {Smail}, {Watson},
  {Hutchison}, {Rigby}, {Welch}, {Ao}, {Bradley}, {Caminha}, {Caputi},
  {Espada}, {Endsley}, {Fudamoto}, {Gonz{\'a}lez-L{\'o}pez}, {Hatsukade},
  {Koekemoer}, {Kokorev}, {Laporte}, {Lee}, {Magdis}, {Ono}, {Rizzo},
  {Shibuya}, {Shimasaku}, {Sun}, {Toft}, {Umehata}, {Wang}, \&
  {Yajima}}]{fujimoto-etal2025}
{Fujimoto}, S., {Ouchi}, M., {Kohno}, K., {et~al.} 2025, Nature Astronomy, 9,
  1553, \dodoi{10.1038/s41550-025-02592-w}

\bibitem[{{Gadotti}(2009)}]{gadotti2009}
{Gadotti}, D.~A. 2009, \mnras, 393, 1531,
  \dodoi{10.1111/j.1365-2966.2008.14257.x}

\bibitem[{{Gao} {et~al.}(2020){Gao}, {Ho}, {Barth}, \& {Li}}]{gao_etal2020}
{Gao}, H., {Ho}, L.~C., {Barth}, A.~J., \& {Li}, Z.-Y. 2020, \apjs, 247, 20,
  \dodoi{10.3847/1538-4365/ab67b2}

\bibitem[{{Genin} {et~al.}(2025){Genin}, {Shuntov}, {Brammer}, {Allen}, {Ito},
  {Magdis}, {Matharu}, {Oesch}, {Toft}, \& {Valentino}}]{genin_etal2025}
{Genin}, A., {Shuntov}, M., {Brammer}, G., {et~al.} 2025, \aap, 699, A343,
  \dodoi{10.1051/0004-6361/202555504}

\bibitem[{{Graham}(2002)}]{graham2002}
{Graham}, A.~W. 2002, \mnras, 334, 859,
  \dodoi{10.1046/j.1365-8711.2002.05548.x}

\bibitem[{{Hainline} {et~al.}(2024){Hainline}, {Johnson}, {Robertson},
  {Tacchella}, {Helton}, {Sun}, {Eisenstein}, {Simmonds}, {Topping}, {Whitler},
  {Willmer}, {Rieke}, {Suess}, {Hviding}, {Cameron}, {Alberts}, {Baker},
  {Baum}, {Bhatawdekar}, {Bonaventura}, {Boyett}, {Bunker}, {Carniani},
  {Charlot}, {Chevallard}, {Chen}, {Curti}, {Curtis-Lake}, {D'Eugenio},
  {Egami}, {Endsley}, {Hausen}, {Ji}, {Looser}, {Lyu}, {Maiolino}, {Nelson},
  {Pusk{\'a}s}, {Rawle}, {Sandles}, {Saxena}, {Smit}, {Stark}, {Williams},
  {Willott}, \& {Witstok}}]{hainline-etal2024}
{Hainline}, K.~N., {Johnson}, B.~D., {Robertson}, B., {et~al.} 2024, \apj, 964,
  71, \dodoi{10.3847/1538-4357/ad1ee4}

\bibitem[{{Hamabe} \& {Kormendy}(1987)}]{Hamabe_Kormendy1987}
{Hamabe}, M., \& {Kormendy}, J. 1987, in IAU Symposium, Vol. 127, Structure and
  Dynamics of Elliptical Galaxies, ed. P.~T. {de Zeeuw}, 379,
  \dodoi{10.1007/978-94-009-3971-4_32}

\bibitem[{{Heintz} {et~al.}(2023){Heintz}, {Brammer}, {Gim{\'e}nez-Arteaga},
  {Strait}, {Lagos}, {Vijayan}, {Matthee}, {Watson}, {Mason}, {Hutter}, {Toft},
  {Fynbo}, \& {Oesch}}]{Heintz-etal2023}
{Heintz}, K.~E., {Brammer}, G.~B., {Gim{\'e}nez-Arteaga}, C., {et~al.} 2023,
  Nature Astronomy, 7, 1517, \dodoi{10.1038/s41550-023-02078-7}

\bibitem[{{Hopkins} {et~al.}(2008){Hopkins}, {Cox}, \&
  {Hernquist}}]{hopkins_etal2008}
{Hopkins}, P.~F., {Cox}, T.~J., \& {Hernquist}, L. 2008, \apj, 689, 17,
  \dodoi{10.1086/592105}

\bibitem[{{Hopkins} {et~al.}(2009{\natexlab{a}}){Hopkins}, {Hernquist}, {Cox},
  {Keres}, \& {Wuyts}}]{hopkins_etal2009a}
{Hopkins}, P.~F., {Hernquist}, L., {Cox}, T.~J., {Keres}, D., \& {Wuyts}, S.
  2009{\natexlab{a}}, \apj, 691, 1424, \dodoi{10.1088/0004-637X/691/2/1424}

\bibitem[{{Hopkins} {et~al.}(2009{\natexlab{b}}){Hopkins}, {Somerville}, {Cox},
  {Hernquist}, {Jogee}, {Kere{\v{s}}}, {Ma}, {Robertson}, \&
  {Stewart}}]{hopkins_etal2009b}
{Hopkins}, P.~F., {Somerville}, R.~S., {Cox}, T.~J., {et~al.}
  2009{\natexlab{b}}, \mnras, 397, 802,
  \dodoi{10.1111/j.1365-2966.2009.14983.x}

\bibitem[{{Hu} {et~al.}(2024){Hu}, {Wang}, {Ge}, {Zhu}, \&
  {Zeng}}]{hu_etal2024}
{Hu}, J., {Wang}, L., {Ge}, J., {Zhu}, K., \& {Zeng}, G. 2024, \mnras, 529,
  4565, \dodoi{10.1093/mnras/stae827}

\bibitem[{{Hu} {et~al.}(2025){Hu}, {Papovich}, {Shen}, {Peng}, {Yung},
  {Lemaux}, {Spilker}, \& {Cole}}]{hu-etal2025}
{Hu}, W., {Papovich}, C., {Shen}, L., {et~al.} 2025, Nature Astronomy, 9, 1568,
  \dodoi{10.1038/s41550-025-02636-1}

\bibitem[{{Huertas-Company} {et~al.}(2024){Huertas-Company}, {Iyer},
  {Angeloudi}, {Bagley}, {Finkelstein}, {Kartaltepe}, {McGrath}, {Sarmiento},
  {Vega-Ferrero}, {Arrabal Haro}, {Behroozi}, {Buitrago}, {Cheng}, {Costantin},
  {Dekel}, {Dickinson}, {Elbaz}, {Grogin}, {Hathi}, {Holwerda}, {Koekemoer},
  {Lucas}, {Papovich}, {P{\'e}rez-Gonz{\'a}lez}, {Pirzkal}, {Seill{\'e}}, {de
  la Vega}, {Wuyts}, {Yang}, \& {Yung}}]{huertas-company-etal2024}
{Huertas-Company}, M., {Iyer}, K.~G., {Angeloudi}, E., {et~al.} 2024, \aap,
  685, A48, \dodoi{10.1051/0004-6361/202346800}

\bibitem[{{JADES}(2023)}]{jades_dataset}
{JADES}. 2023, Data from the JWST Advanced Deep Extragalactic Survey (JADES),
  STScI/MAST, \dodoi{10.17909/8TDJ-8N28}

\bibitem[{{Ji} {et~al.}(2024){Ji}, {Williams}, {Tacchella}, {Suess}, {Baker},
  {Alberts}, {Bunker}, {Johnson}, {Robertson}, {Sun}, {Eisenstein}, {Rieke},
  {Maseda}, {Hainline}, {Hausen}, {Rieke}, {Willmer}, {Egami}, {Shivaei},
  {Carniani}, {Charlot}, {Chevallard}, {Curtis-Lake}, {Looser}, {Maiolino},
  {Willott}, {Chen}, {Helton}, {Lyu}, {Nelson}, {Bhatawdekar}, {Boyett}, \&
  {Sandles}}]{ji-etal2024}
{Ji}, Z., {Williams}, C.~C., {Tacchella}, S., {et~al.} 2024, \apj, 974, 135,
  \dodoi{10.3847/1538-4357/ad6e7f}

\bibitem[{{Kartaltepe} {et~al.}(2023){Kartaltepe}, {Rose}, {Vanderhoof},
  {McGrath}, {Costantin}, {Cox}, {Yung}, {Kocevski}, {Wuyts}, {Ferguson},
  {Bagley}, {Finkelstein}, {Amor{\'\i}n}, {Andrews}, {Arrabal Haro},
  {Backhaus}, {Behroozi}, {Bisigello}, {Calabr{\`o}}, {Casey}, {Coogan},
  {Cooper}, {Croton}, {de la Vega}, {Dickinson}, {Fontana}, {Franco},
  {Grazian}, {Grogin}, {Hathi}, {Holwerda}, {Huertas-Company}, {Iyer}, {Jogee},
  {Jung}, {Kewley}, {Kirkpatrick}, {Koekemoer}, {Liu}, {Lotz}, {Lucas},
  {Newman}, {Pacifici}, {Pandya}, {Papovich}, {Pentericci},
  {P{\'e}rez-Gonz{\'a}lez}, {Petersen}, {Pirzkal}, {Rafelski}, {Ravindranath},
  {Simons}, {Snyder}, {Somerville}, {Stanway}, {Straughn}, {Tacchella},
  {Trump}, {Vega-Ferrero}, {Wilkins}, {Yang}, \&
  {Zavala}}]{kartaltepe-etal2023}
{Kartaltepe}, J.~S., {Rose}, C., {Vanderhoof}, B.~N., {et~al.} 2023, \apjl,
  946, L15, \dodoi{10.3847/2041-8213/acad01}

\bibitem[{{Khanday} {et~al.}(2022){Khanday}, {Saha}, {Iqbal}, {Dhiwar}, \&
  {Pahwa}}]{khanday_etal2022}
{Khanday}, S.~A., {Saha}, K., {Iqbal}, N., {Dhiwar}, S., \& {Pahwa}, I. 2022,
  \mnras, 515, 5043, \dodoi{10.1093/mnras/stac2009}

\bibitem[{{Kormendy}(1977)}]{Kormendy1977}
{Kormendy}, J. 1977, \apj, 218, 333, \dodoi{10.1086/155687}

\bibitem[{{Krajnovi{\'c}} {et~al.}(2013){Krajnovi{\'c}}, {Alatalo}, {Blitz},
  {Bois}, {Bournaud}, {Bureau}, {Cappellari}, {Davies}, {Davis}, {de Zeeuw},
  {Duc}, {Emsellem}, {Khochfar}, {Kuntschner}, {McDermid}, {Morganti}, {Naab},
  {Oosterloo}, {Sarzi}, {Scott}, {Serra}, {Weijmans}, \&
  {Young}}]{Krajnovic_etal2013}
{Krajnovi{\'c}}, D., {Alatalo}, K., {Blitz}, L., {et~al.} 2013, \mnras, 432,
  1768, \dodoi{10.1093/mnras/sts315}

\bibitem[{{La Barbera} {et~al.}(2003){La Barbera}, {Busarello}, {Merluzzi},
  {Massarotti}, \& {Capaccioli}}]{LaBarbera_etal2003}
{La Barbera}, F., {Busarello}, G., {Merluzzi}, P., {Massarotti}, M., \&
  {Capaccioli}, M. 2003, \apj, 595, 127, \dodoi{10.1086/377250}

\bibitem[{{La Barbera} {et~al.}(2010){La Barbera}, {de Carvalho}, {de La Rosa},
  \& {Lopes}}]{LaBarbera_etal2010}
{La Barbera}, F., {de Carvalho}, R.~R., {de La Rosa}, I.~G., \& {Lopes},
  P.~A.~A. 2010, \mnras, 408, 1335, \dodoi{10.1111/j.1365-2966.2010.17091.x}

\bibitem[{{Longhetti} {et~al.}(2007){Longhetti}, {Saracco}, {Severgnini},
  {Della Ceca}, {Mannucci}, {Bender}, {Drory}, {Feulner}, \&
  {Hopp}}]{longhetti_etal2007}
{Longhetti}, M., {Saracco}, P., {Severgnini}, P., {et~al.} 2007, \mnras, 374,
  614, \dodoi{10.1111/j.1365-2966.2006.11171.x}

\bibitem[{{Merlin} {et~al.}(2012){Merlin}, {Chiosi}, {Piovan}, {Grassi},
  {Buonomo}, \& {La Barbera}}]{merlin_etal2012}
{Merlin}, E., {Chiosi}, C., {Piovan}, L., {et~al.} 2012, \mnras, 427, 1530,
  \dodoi{10.1111/j.1365-2966.2012.21965.x}

\bibitem[{{Meyer} {et~al.}(2024){Meyer}, {Oesch}, {Giovinazzo}, {Weibel},
  {Brammer}, {Matthee}, {Naidu}, {Bouwens}, {Chisholm}, {Covelo-Paz},
  {Fudamoto}, {Maseda}, {Nelson}, {Shivaei}, {Xiao}, {Herard-Demanche},
  {Illingworth}, {Kerutt}, {Kramarenko}, {Labbe}, {Leonova}, {Magee},
  {Matharu}, {Prieto Lyon}, {Reddy}, {Schaerer}, {Shapley}, {Stefanon},
  {Wozniak}, \& {Wuyts}}]{meyer-etal2024}
{Meyer}, R.~A., {Oesch}, P.~A., {Giovinazzo}, E., {et~al.} 2024, \mnras, 535,
  1067, \dodoi{10.1093/mnras/stae2353}

\bibitem[{{Morishita} {et~al.}(2024){Morishita}, {Stiavelli}, {Chary},
  {Trenti}, {Bergamini}, {Chiaberge}, {Leethochawalit}, {Roberts-Borsani},
  {Shen}, \& {Treu}}]{morishita-etal2024}
{Morishita}, T., {Stiavelli}, M., {Chary}, R.-R., {et~al.} 2024, \apj, 963, 9,
  \dodoi{10.3847/1538-4357/ad1404}

\bibitem[{{Naab} {et~al.}(2009){Naab}, {Johansson}, \&
  {Ostriker}}]{naab_etal2009}
{Naab}, T., {Johansson}, P.~H., \& {Ostriker}, J.~P. 2009, \apjl, 699, L178,
  \dodoi{10.1088/0004-637X/699/2/L178}

\bibitem[{{Naab} {et~al.}(2007){Naab}, {Johansson}, {Ostriker}, \&
  {Efstathiou}}]{naab_etal2007}
{Naab}, T., {Johansson}, P.~H., {Ostriker}, J.~P., \& {Efstathiou}, G. 2007,
  \apj, 658, 710, \dodoi{10.1086/510841}

\bibitem[{{Nigoche-Netro} {et~al.}(2008){Nigoche-Netro}, {Ruelas-Mayorga}, \&
  {Franco-Balderas}}]{Nigoche-Netro_etal2008}
{Nigoche-Netro}, A., {Ruelas-Mayorga}, A., \& {Franco-Balderas}, A. 2008, \aap,
  491, 731, \dodoi{10.1051/0004-6361:200810211}

\bibitem[{{Nipoti} {et~al.}(2003){Nipoti}, {Londrillo}, \&
  {Ciotti}}]{nipoti_etal2003}
{Nipoti}, C., {Londrillo}, P., \& {Ciotti}, L. 2003, \mnras, 342, 501,
  \dodoi{10.1046/j.1365-8711.2003.06554.x}

\bibitem[{{Oesch} {et~al.}(2023){Oesch}, {Brammer}, {Naidu}, {Bouwens},
  {Chisholm}, {Illingworth}, {Matthee}, {Nelson}, {Qin}, {Reddy}, {Shapley},
  {Shivaei}, {van Dokkum}, {Weibel}, {Whitaker}, {Wuyts}, {Covelo-Paz},
  {Endsley}, {Fudamoto}, {Giovinazzo}, {Herard-Demanche}, {Kerutt},
  {Kramarenko}, {Labbe}, {Leonova}, {Lin}, {Magee}, {Marchesini}, {Maseda},
  {Mason}, {Matharu}, {Meyer}, {Neufeld}, {Prieto Lyon}, {Schaerer}, {Sharma},
  {Shuntov}, {Smit}, {Stefanon}, {Wyithe}, \& {Xiao}}]{oesch-etal2023}
{Oesch}, P.~A., {Brammer}, G., {Naidu}, R.~P., {et~al.} 2023, \mnras, 525,
  2864, \dodoi{10.1093/mnras/stad2411}

\bibitem[{{Oke}(1974)}]{oke1974}
{Oke}, J.~B. 1974, \apjs, 27, 21, \dodoi{10.1086/190287}

\bibitem[{{Oser} {et~al.}(2010){Oser}, {Ostriker}, {Naab}, {Johansson}, \&
  {Burkert}}]{oser_etal2010}
{Oser}, L., {Ostriker}, J.~P., {Naab}, T., {Johansson}, P.~H., \& {Burkert}, A.
  2010, \apj, 725, 2312, \dodoi{10.1088/0004-637X/725/2/2312}

\bibitem[{{Park} {et~al.}(2022){Park}, {Lee}, {Kim}, {Jeong}, {Pichon},
  {Gibson}, {Snaith}, {Shin}, {Kim}, {Dubois}, \& {Few}}]{park-etal2022}
{Park}, C., {Lee}, J., {Kim}, J., {et~al.} 2022, \apj, 937, 15,
  \dodoi{10.3847/1538-4357/ac85b5}

\bibitem[{{Pastrav}(2021)}]{pastrav2021}
{Pastrav}, B.~A. 2021, \mnras, 506, 452, \dodoi{10.1093/mnras/stab1746}

\bibitem[{{Peng} {et~al.}(2002){Peng}, {Ho}, {Impey}, \& {Rix}}]{peng_etal2002}
{Peng}, C.~Y., {Ho}, L.~C., {Impey}, C.~D., \& {Rix}, H.-W. 2002, \aj, 124,
  266, \dodoi{10.1086/340952}

\bibitem[{{Peng} {et~al.}(2010){Peng}, {Ho}, {Impey}, \& {Rix}}]{peng_etal2010}
---. 2010, \aj, 139, 2097, \dodoi{10.1088/0004-6256/139/6/2097}

\bibitem[{{Pollock} {et~al.}(2026){Pollock}, {Heintz}, {Witstok},
  {Gottumukkala}, {Brammer}, {Bose}, {Cameron}, {Dayal}, {van Dokkum}, {Fynbo},
  {Gelli}, {Hayes}, {Inoue}, {Lagos}, {Laursen}, {Meyer}, {Naidu}, {Oesch},
  {Rowland}, {Tanvir}, {Tacchella}, {Terp}, {Valentino}, {Walter}, {Weaver}, \&
  {Witten}}]{Pollock-etal2026}
{Pollock}, C.~L., {Heintz}, K.~E., {Witstok}, J., {et~al.} 2026, arXiv
  e-prints, arXiv:2602.11783, \dodoi{10.48550/arXiv.2602.11783}

\bibitem[{{Prugniel} \& {Simien}(1996)}]{Prugniel_Simien1996}
{Prugniel}, P., \& {Simien}, F. 1996, \aap, 309, 749

\bibitem[{{Pusk{\'a}s} {et~al.}(2025){Pusk{\'a}s}, {Tacchella}, {Simmonds},
  {Hainline}, {D'Eugenio}, {Alberts}, {Arribas}, {Baker}, {Bunker}, {Carniani},
  {Charlot}, {Duan}, {Eisenstein}, {Ji}, {Johnson}, {Jones}, {Maiolino},
  {McClymont}, {Rieke}, {Rinaldi}, {Robertson}, {{\"U}bler}, {Williams},
  {Willmer}, {Willott}, \& {Witstok}}]{puskas-etal2025}
{Pusk{\'a}s}, D., {Tacchella}, S., {Simmonds}, C., {et~al.} 2025, \mnras, 540,
  2146, \dodoi{10.1093/mnras/staf813}

\bibitem[{{Reda} {et~al.}(2005){Reda}, {Forbes}, \& {Hau}}]{reda_etal2005}
{Reda}, F.~M., {Forbes}, D.~A., \& {Hau}, G. K.~T. 2005, \mnras, 360, 693,
  \dodoi{10.1111/j.1365-2966.2005.09058.x}

\bibitem[{{Rettura} {et~al.}(2010){Rettura}, {Rosati}, {Nonino}, {Fosbury},
  {Gobat}, {Menci}, {Strazzullo}, {Mei}, {Demarco}, \&
  {Ford}}]{Rettura_etal2010}
{Rettura}, A., {Rosati}, P., {Nonino}, M., {et~al.} 2010, \apj, 709, 512,
  \dodoi{10.1088/0004-637X/709/1/512}

\bibitem[{{Rieke} {et~al.}(2023){Rieke}, {Robertson}, {Tacchella}, {Hainline},
  {Johnson}, {Hausen}, {Ji}, {Willmer}, {Eisenstein}, {Pusk{\'a}s}, {Alberts},
  {Arribas}, {Baker}, {Baum}, {Bhatawdekar}, {Bonaventura}, {Boyett}, {Bunker},
  {Cameron}, {Carniani}, {Charlot}, {Chevallard}, {Chen}, {Curti},
  {Curtis-Lake}, {Danhaive}, {DeCoursey}, {Dressler}, {Egami}, {Endsley},
  {Helton}, {Hviding}, {Kumari}, {Looser}, {Lyu}, {Maiolino}, {Maseda},
  {Nelson}, {Rieke}, {Rix}, {Sandles}, {Saxena}, {Sharpe}, {Shivaei},
  {Skarbinski}, {Smit}, {Stark}, {Stone}, {Suess}, {Sun}, {Topping},
  {{\"U}bler}, {Villanueva}, {Wallace}, {Williams}, {Willott}, {Whitler},
  {Witstok}, \& {Woodrum}}]{rieke-etal2023}
{Rieke}, M.~J., {Robertson}, B., {Tacchella}, S., {et~al.} 2023, \apjs, 269,
  16, \dodoi{10.3847/1538-4365/acf44d}

\bibitem[{{Robertson} {et~al.}(2006){Robertson}, {Cox}, {Hernquist}, {Franx},
  {Hopkins}, {Martini}, \& {Springel}}]{robertson_etal2006}
{Robertson}, B., {Cox}, T.~J., {Hernquist}, L., {et~al.} 2006, \apj, 641, 21,
  \dodoi{10.1086/500360}

\bibitem[{{Sachdeva} {et~al.}(2020){Sachdeva}, {Ho}, {Li}, \&
  {Shankar}}]{sachdeva_etal2020}
{Sachdeva}, S., {Ho}, L.~C., {Li}, Y.~A., \& {Shankar}, F. 2020, \apj, 899, 89,
  \dodoi{10.3847/1538-4357/aba82d}

\bibitem[{{Sachdeva} \& {Saha}(2018)}]{sachdeva_saha2018}
{Sachdeva}, S., \& {Saha}, K. 2018, \mnras, 478, 41,
  \dodoi{10.1093/mnras/sty1084}

\bibitem[{{Sachdeva} {et~al.}(2017){Sachdeva}, {Saha}, \&
  {Singh}}]{sachdeva_etal2017}
{Sachdeva}, S., {Saha}, K., \& {Singh}, H.~P. 2017, \apj, 840, 79,
  \dodoi{10.3847/1538-4357/aa6c61}

\bibitem[{{Saracco} {et~al.}(2017){Saracco}, {Gargiulo}, {Ciocca}, \&
  {Marchesini}}]{saracco_etal2017}
{Saracco}, P., {Gargiulo}, A., {Ciocca}, F., \& {Marchesini}, D. 2017, \aap,
  597, A122, \dodoi{10.1051/0004-6361/201628866}

\bibitem[{{Saracco} {et~al.}(2010){Saracco}, {Longhetti}, \&
  {Gargiulo}}]{saracco_etal2010}
{Saracco}, P., {Longhetti}, M., \& {Gargiulo}, A. 2010, \mnras, 408, L21,
  \dodoi{10.1111/j.1745-3933.2010.00920.x}

\bibitem[{{Saracco} {et~al.}(2014){Saracco}, {Casati}, {Gargiulo}, {Longhetti},
  {Lonoce}, {Tamburri}, {Bettoni}, {D'Onofrio D'Onofrio}, {Fasano},
  {Poggianti}, {Boutsia}, {Fumana}, \& {Sani}}]{saracco_etal2014}
{Saracco}, P., {Casati}, A., {Gargiulo}, A., {et~al.} 2014, \aap, 567, A94,
  \dodoi{10.1051/0004-6361/201423495}

\bibitem[{{Schlafly} \& {Finkbeiner}(2011)}]{SchlaflyFinkbeiner2011}
{Schlafly}, E.~F., \& {Finkbeiner}, D.~P. 2011, \apj, 737, 103,
  \dodoi{10.1088/0004-637X/737/2/103}

\bibitem[{{Schlegel} {et~al.}(1998){Schlegel}, {Finkbeiner}, \&
  {Davis}}]{Schlegel-etal1998}
{Schlegel}, D.~J., {Finkbeiner}, D.~P., \& {Davis}, M. 1998, \apj, 500, 525,
  \dodoi{10.1086/305772}

\bibitem[{{Steinmetz} \& {Navarro}(1999)}]{Steinmetz_Navarro1999}
{Steinmetz}, M., \& {Navarro}, J.~F. 1999, \apj, 513, 555,
  \dodoi{10.1086/306904}

\bibitem[{{Tacchella} {et~al.}(2023){Tacchella}, {Johnson}, {Robertson},
  {Carniani}, {D'Eugenio}, {Kumari}, {Maiolino}, {Nelson}, {Suess},
  {{\"U}bler}, {Williams}, {Adebusola}, {Alberts}, {Arribas}, {Bhatawdekar},
  {Bonaventura}, {Bowler}, {Bunker}, {Cameron}, {Curti}, {Egami}, {Eisenstein},
  {Frye}, {Hainline}, {Helton}, {Ji}, {Looser}, {Lyu}, {Perna}, {Rawle},
  {Rieke}, {Rieke}, {Saxena}, {Sandles}, {Shivaei}, {Simmonds}, {Sun},
  {Willmer}, {Willott}, \& {Witstok}}]{Tacchella-etal2023}
{Tacchella}, S., {Johnson}, B.~D., {Robertson}, B.~E., {et~al.} 2023, \mnras,
  522, 6236, \dodoi{10.1093/mnras/stad1408}

\bibitem[{{Tacconi} {et~al.}(2020){Tacconi}, {Genzel}, \&
  {Sternberg}}]{Tacconi-etal2020}
{Tacconi}, L.~J., {Genzel}, R., \& {Sternberg}, A. 2020, \araa, 58, 157,
  \dodoi{10.1146/annurev-astro-082812-141034}

\bibitem[{{Tacconi} {et~al.}(2010){Tacconi}, {Genzel}, {Neri}, {Cox}, {Cooper},
  {Shapiro}, {Bolatto}, {Bouch{\'e}}, {Bournaud}, {Burkert}, {Combes},
  {Comerford}, {Davis}, {F{\"o}rster Schreiber}, {Garcia-Burillo},
  {Gracia-Carpio}, {Lutz}, {Naab}, {Omont}, {Shapley}, {Sternberg}, \&
  {Weiner}}]{tacconi_etal2010}
{Tacconi}, L.~J., {Genzel}, R., {Neri}, R., {et~al.} 2010, \nat, 463, 781,
  \dodoi{10.1038/nature08773}

\bibitem[{{Tortorelli} {et~al.}(2018){Tortorelli}, {Mercurio}, {Paolillo},
  {Rosati}, {Gargiulo}, {Gobat}, {Balestra}, {Caminha}, {Annunziatella},
  {Grillo}, {Lombardi}, {Nonino}, {Rettura}, {Sartoris}, \&
  {Strazzullo}}]{Tortorelli_etal2018}
{Tortorelli}, L., {Mercurio}, A., {Paolillo}, M., {et~al.} 2018, \mnras, 477,
  648, \dodoi{10.1093/mnras/sty617}

\bibitem[{{Tortorelli} {et~al.}(2023){Tortorelli}, {Mercurio}, {Granata},
  {Rosati}, {Grillo}, {Nonino}, {Acebron}, {Angora}, {Bergamini}, {Caminha},
  {Me{\v{s}}tri{\'c}}, \& {Vanzella}}]{Tortorelli_etal2023}
{Tortorelli}, L., {Mercurio}, A., {Granata}, G., {et~al.} 2023, \aap, 671, L9,
  \dodoi{10.1051/0004-6361/202346151}

\bibitem[{{Tully} \& {Fisher}(1977)}]{TullyFisher1977}
{Tully}, R.~B., \& {Fisher}, J.~R. 1977, \aap, 54, 661

\bibitem[{{van der Wel} {et~al.}(2008){van der Wel}, {Holden}, {Zirm}, {Franx},
  {Rettura}, {Illingworth}, \& {Ford}}]{van_der_Wel_etal2008}
{van der Wel}, A., {Holden}, B.~P., {Zirm}, A.~W., {et~al.} 2008, \apj, 688,
  48, \dodoi{10.1086/592267}

\bibitem[{{Ziegler} {et~al.}(1999){Ziegler}, {Saglia}, {Bender}, {Belloni},
  {Greggio}, \& {Seitz}}]{Ziegler_etal1999}
{Ziegler}, B.~L., {Saglia}, R.~P., {Bender}, R., {et~al.} 1999, \aap, 346, 13,
  \dodoi{10.48550/arXiv.astro-ph/9903222}

\bibitem[{{Z{\"o}ller} {et~al.}(2024){Z{\"o}ller}, {Kluge}, {Staiger}, \&
  {Bender}}]{zoller-etal2024}
{Z{\"o}ller}, R., {Kluge}, M., {Staiger}, B., \& {Bender}, R. 2024, \apjs, 271,
  52, \dodoi{10.3847/1538-4365/ad2775}

\bibitem[{{Zolotov} {et~al.}(2015){Zolotov}, {Dekel}, {Mandelker}, {Tweed},
  {Inoue}, {DeGraf}, {Ceverino}, {Primack}, {Barro}, \&
  {Faber}}]{zolotov-etal2015}
{Zolotov}, A., {Dekel}, A., {Mandelker}, N., {et~al.} 2015, \mnras, 450, 2327,
  \dodoi{10.1093/mnras/stv740}

\end{thebibliography}
\bibliographystyle{aasjournal}

\appendix

\section{Effect of \textsc{GALFIT} parameter bounds on the recovered structural parameters}
\label{app:constraints}

To assess the robustness of the structural parameters used in this work, we compare the results obtained from \textsc{GALFIT} with and without the adopted parameter bounds. As discussed in section 2, conservative limits were imposed on the fitting parameters to avoid unphysical solutions and improve numerical stability when modelling compact, high-\textit{z} galaxies. Here, we examine whether these bounds introduce systematic biases in the recovered structural properties.

\begin{figure*}
\centering
\includegraphics[width=\textwidth]{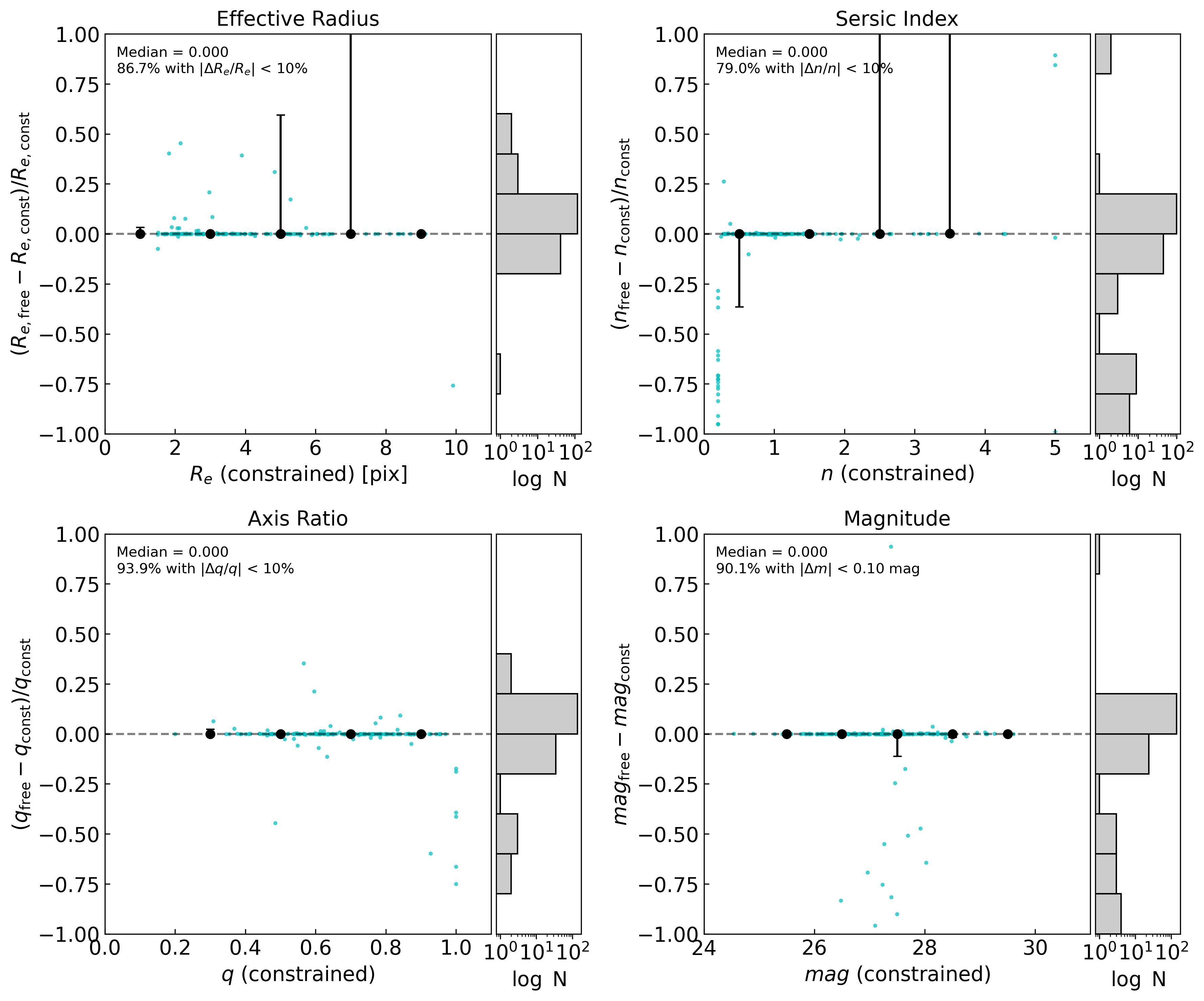}
\caption{
\textbf{Comparison between structural parameters obtained from unconstrained and constrained \textsc{GALFIT} modelling}. The panels show the fractional differences in effective radius ($\rm R_{e}$), Sersic index ($n$), and axis ratio ($q$), together with the absolute magnitude difference, as a function of the corresponding parameter derived from the constrained fits. Cyan points represent individual galaxies, while the black symbols indicate binned median values. The histograms on the right show the distributions of the parameter differences. The percentages reported in each panel indicate the fraction of galaxies for which the constrained and unconstrained measurements agree within 10\% (or 0.1 mag for the magnitude). The errorbars represent 16th and 84th percentile errors on the median.
}
\label{fig:constraint_test}
\end{figure*}

To understand the impact of the constraints, we repeated the fitting for the galaxies in our sample, modelled with a single-sersic component, without imposing the \textsc{GALFIT} parameter bounds and compared the recovered parameters with those obtained from the constrained fits. Figure~\ref{fig:constraint_test} presents the fractional differences between the unconstrained and constrained measurements of effective radius (R$_{\rm e}$), Sersic index ($n$), axis ratio ($q$), and magnitude.

Overall, the constrained and unconstrained solutions are in good agreement. The distributions of fractional differences in the parameters (simple difference in case of magnitudes) are centered around zero, with median offsets consistent with zero for all quantities. We find that 86.7\% of galaxies have $|\Delta \rm R_{e}/R_{e}| < 10\%$, 79.0\% have $|\Delta n/n| < 10\%$, 93.9\% have $|\Delta q/q| < 10\%$, and 90.1\% have magnitude differences smaller than 0.1 mag. For brevity, we have not extended the axes ranges beyond $\pm$1 as they are unphysical for galaxies at these redshifts. The small fraction of larger deviations occurs predominantly for sources with unusually large \textit{n} values, e.g., $\sim19.5$. This results in extreme solutions for R$_{\rm e}$, e.g., 9.2 kpc at \textit{z=7}. Importantly, these cases do not dominate the sample and do not affect the overall distributions of structural parameters.

This comparison demonstrates that the adopted constraints do not introduce significant systematic biases into the recovered structural parameters. Rather, they serve to stabilize the fitting procedure while preserving the underlying galaxy properties. We therefore conclude that the structural measurements derived throughout this work, and consequently the derived KR, are robust against the specific choice of fitting constraints.

\section{Effect of excluding DJA galaxies on the KR}
\label{app:dja}

\begin{figure*}
\centering
\includegraphics[width=\textwidth]{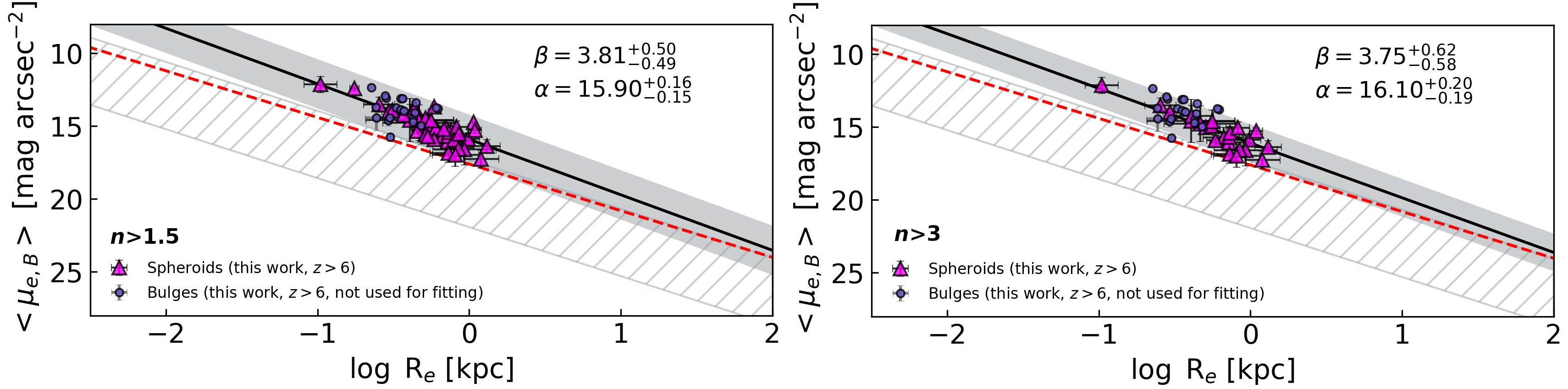}
\caption{
\textbf{Kormendy relations derived after excluding all DJA-catalog galaxies from the Samples 1 and 2.} The left and right panels show the results for the $n>1.5$ and $n>3$ samples, respectively. All markers, curves and shaded regions follow the same conventions as in Figure~\ref{fig:KR-combined}.
}
\label{fig:KR_no_DJA}
\end{figure*}

As discussed in Section~4, the structural parameters adopted from the DJA catalog were based on multiband modelling and represent weighted structural parameters derived over the wavelength range $0.8-5~\mu$m. The KR fits after excluding the DJA galaxies are shown in Figure~\ref{fig:KR_no_DJA}. We do not observe any significant and systematic shift in either the slope or zero-point after excluding the DJA galaxies. The primary effect of excluding the DJA galaxies is a mild increase in the uncertainties of these fitted parameters. However, the intrinsic scatter decreases from 0.80 to 0.56 and from 0.98 to 0.43 for the $n>1.5$ and $n>3$ samples, respectively. This behaviour suggests the possibility that the DJA galaxies are contributing to an additional diversity of the spheroidal galaxy population in the KR plane. However, we cannot rule out the possibility that the differences in the fitting methodologies have contributed to this change in the scatter. Given the limited sample size of the DJA galaxies, we refrain from drawing strong conclusions regarding the origin of this difference. Larger homogeneous samples of $z\gtrsim6$ galaxies will be required to determine whether the observed reduction in scatter reflects a genuine population difference or arises from sample variance. This demonstrates that the inferred KR slope and zero-point, and hence the main conclusions drawn in this work are not significantly affected by the inclusion of galaxies whose structural parameters were derived using a different fitting methodology.

\end{document}